\documentclass[
showpacs,preprintnumbers,
amsmath,amssymb,
aps,
twocolumn
]{revtex4}

\usepackage{graphicx}
\usepackage{dcolumn}
\usepackage{bm}
\usepackage[dvipdfm, usenames]{color}
\usepackage{here}

\def\br{\begin{align}}
\def\er{\end{align}}
\def\be{\begin{equation}}
\def\ee{\end{equation}}
\def\({\left(}
\def\){\right)}

\def\rlx{\relax\leavevmode}
\def\IR{\rlx\hbox{\rm I\kern-.18em R}}

\def\u2{\mid u\mid^2}

\def\IZ{\rlx\hbox{\sf Z\kern-.4em Z}}
\def\IR{\rlx\hbox{\rm I\kern-.18em R}}
\def\IC{\rlx\hbox{\,$\inbar\kern-.3em{\rm C}$}}

\begin{document}

\preprint{APS/123-QED}

\title{Exact, molecular-shaped vortices with fractional and integer charges in the extended Skyrme-Faddeev model}%

\author{Nobuyuki Sawado}
\email{sawado(at)ph.noda.tus.ac.jp}
\author{Yuta Tamaki}
\email{mojyamojya.sax.0313@gmail.com}

\affiliation{Department of Physics, Faculty of Science and Technology,
Tokyo University of Science, Noda, Chiba 278-8510, Japan}

\date{\today}

\begin{abstract}
We analytically construct vortex solutions in the integrable sector of the extended Skyrme-Faddeev model.
The solutions are holomorphic type which satisfy the zero curvature condition. 
For the model parameter $\beta e^2=1$ there is a lump solution, and 
for $\beta e^2 \neq 1$ new potentials are introduced for the several molecular-shaped solutions with half-integer or integer charges.
They necessarily have infinite number of conserved currents and some of the examples are shown. 
By performing an annealing simulation with our potentials, we verify the existence of the solutions of the integrable sector.   
\end{abstract}


\maketitle


\section{Introduction}

The Skyrme-Faddeev model was introduced in the seventies \cite{sf} as a clever generalization to $(3+1)$ dimensions of the $O(3)$ non-linear sigma model in
$(2+1)$ dimensions \cite{bp}. The Skyrme term, quartic in derivatives of the field, balances the quadratic kinetic term and according to Derrick's theorem, allows the@existence of stable solutions with non-trivial Hopf topological charges. Due to the highly non-linear character of the model and the lack of symmetries, the first soliton solutions were only constructed in the late nineties using numerical methods
\cite{Gladikowski:1996mb,solfn,sutcliffe,hietarinta}. 
Several physical applications based on the model have been extensively studied in many areas due mainly to the knotted character of the solutions~\cite{babaev}. 
The numerical efforts in the construction of the solutions have improved our understanding of the properties of the model \cite{improve} and even the scattering of knotted solitons has been investigated
\cite{hietarinta-scatter}. One of the aspects of the model that has attracted considerable attention has been its connection with gauge theories. Faddeev and Niemi have conjectured that it might describe the low energy limit of the pure $SU(2)$ Yang-Mills theory
\cite{fn}. They based their argument on a decomposition of the physical degrees of freedom
of the $SU(2)$ connection, proposed in the eighties by Cho \cite{chofn}, and involving a
triplet of scalar fields ${\vec n}$ taking values on the sphere $S^2$ (${\vec n}^2=1$).
Gies \cite {gies} has calculated the Wilsonian one loop effective action for the pure $SU(2)$ 
Yang-Mills theory assuming
Cho's decomposition, and found that the Skyrme-Faddeev action is indeed part of it, but
additional quartic terms in the derivatives of the triplet ${\vec n}$ are unavoidable. 

The extended version of such Skyrme-Faddeev (ESF) model,
\begin{align}
\mathcal{L}= \mathcal{M}^2 \partial_{\mu}\vec{n} \cdot \partial^{\mu}\vec{n} 
- \frac{1}{e^{2}}(\partial_{\mu}\vec{n} \wedge \partial_{\nu} \vec{n})^2 \nonumber \\
+ \frac{\beta}{2}(\partial_{\mu}\vec{n} \cdot \partial^{\mu} \vec{n})^2 - V(n_1,n_2,n_3)
\label{lagrangian}
\end{align}
has already been studied: 
The static energy density (${\cal H}_{{\rm static}}=-{\cal L}$) associated to (\ref{lagrangian})
is positive definite if $V>0$, $\mathcal{M}^2>0$, $e^2>0$ and $\beta<0$. That is the sector explored
in \cite{Gladikowski:1996mb} where Hopf soliton solutions were first constructed (for $V=0$). In
addition, that is also the sector explored in \cite{Sawado:2005wa} but with additional
terms involving second derivatives of the ${\vec n}$ field, where Hopf solitons were also constructed. 
The static energy density of (\ref{lagrangian}) is also positive definite for
$V>0$ if
\begin{align}
\mathcal{M}^2>0\,; \qquad  e^2<0\, ; \qquad \beta <0 \, ; \qquad  \beta\, e^2\geq 1.
\label{nicesector}
\end{align}
That is the sector that agrees with the signature of the terms in the one loop effective
action calculated in \cite{gies} and it is the sector that we will consider in this
paper. Static Hopf solitons with the axial symmetry were constructed in \cite{Ferreira:2009gj,Ferreira:2010zz} 
for the sector (\ref{nicesector}) and their quantum excitations, including comparison
with glue ball spectrum, were considered in \cite{quantumhopfions}.  An interesting feature
of the Hopf solitons constructed in \cite{Ferreira:2009gj} is that they shrink in size and
then disappear as $\beta\,e^2\rightarrow 1$. 
Full numerical simulation was followed for  
the existence of such knotted solutions in~\cite{Foster:2012wx}. 

The action (\ref{lagrangian}) also possesses the vortex solutions. 
The first exact vortex solutions for the theory were constructed in \cite{vortexlaf},
 by exploring the integrability properties of a submodel of (\ref{lagrangian}). 
In order to describe those exact vortex solutions, it is better to perform the 
stereographic projection of the target space
$S^2$ onto the plane parameterized by the complex scalar field $u$ and related to 
${\vec n}$ by
\begin{align}
{\vec n} = \(u+u^*,-i\(u-u^*\),\u2 -1\)/\(1+\u2\).
\label{udef}
\end{align}
It was shown in \cite{vortexlaf} that the field configurations of the form 
\begin{align}
u\equiv u\(z,y\),~~u^*\equiv u^*\( z^*,y\)~~~~{\rm for}~~~~ 
\beta\,e^2=1,~~V=0 \nonumber \\
\label{exactclass}
\end{align}
are exact solutions of (\ref{lagrangian}), where $z=x^1+i\,\varepsilon_1\,x^2$ and
$y=x^3-\varepsilon_2\,x^0$, with $\varepsilon_a=\pm 1$, $a=1,2$ (the signs can be 
chosen independently). The $x^{\mu}$,
$\mu=0,1,2,3$, are the Cartesian coordinates of the Minkowski space-time. 
Despite the fact that (\ref{exactclass}) constitutes a very large class of solutions, no finite energy
solutions were found within it. If the dependence of the $u$ field upon the variable $y$
is in the form of phases like $e^{i\,k\,y}$, then one finds solutions with finite energy
per unit of length along the $x^3$-axis. The simplest solution is of the form
$u=z^n\,e^{i\,k\,y}$, with $n$ to be integer, and it corresponds to a vortex parallel to the
$x^3$-axis and with waves travelling along it with the speed of light. More general
solutions of the class (\ref{exactclass}) were constructed in \cite{newsf}, including
multi-vortices separated from each other and all parallel to the $x^3$-axis.

The vortex solutions for the model continue to exist when the condition $\beta e^2=1$ is relaxed
by introducing a potential $V$~\cite{Ferreira:2011mz}.
The potential is essentially introduced to stabilize the vortex solutions. 
It is well known that there are some variations for the potential when it is a functional of the 
third component $n_3$ of the triplet ${\vec n}$, such as so-called {\it old}-baby potential
(one-vacuum type; which has zero at the spatial infinity)~\cite{Leese:1989gi}
 and {\it new}-baby potential (two-vacuum type; which possesses two zeros at the 
origin and the infinity)~\cite{Kudryavtsev:1997nw}. 
The baby-Skyrme model is a 2+1 mimic of the Skyrme model and has 
the static planar solution called baby-skyrmions. The present model has 
close relation with the baby-Skyrme model when we restrict our analysis  
only in the static planar solution.  
In the {\it old}-baby potential, the rotational symmetry of the baby-skyrmions 
is spontaneously broken while in the {\it new}-baby, no such transition of the structure 
occurs~\cite{Hen:2007in}. 
Similar behavior has also been observed in our 
vortex solution~\cite{Ferreira:2012cy}. By using the {\it old}-baby potential, 
the deformation grows as $\beta e^2$ increases.   
Note that if the potential is a functional of the third component $n_3$
it breaks the $O(3)$ symmetry of the original Skyrme-Faddeev down to $O(2)$, 
the group of rotations on the plane $n_1-n_2$, and so eliminating two of the
three Goldstone boson degrees of freedom. 

If one allows to include all components $n_1,n_2$ and $n_3$ in the potential, 
there must be many possibilities of the choice. 
In the easy plane potential $V=\frac{1}{2}m^2 n_3^2$, the baby-skyrmions possess 
the dihedral symmetry $D_2$~\cite{Jaykka:2010bq}. 
Originally, such a symmetrical solution was found by the 
choice of $V=\frac{1}{2}m^2(1-n_3^2)(1-n_1^2)$~\cite{Ward:2003xv}. 
A sophisticated form of the potential $V=m^2 |1-(n_1+in_2)^N|^2(1-n_3)~(N\geqq 2)$ 
is essentially the same as the {\it old}-baby but exhibits the $D_N$ symmetry~\cite{Jaykka:2011ic}. 
The baby-skyrmions or the fractional vortex states for a variant of these potentials 
are extensively investigated in~\cite{Nitta}. 
In general, one can get solutions with dihedral 
symmetries if one adopt potentials which contain terms with 
$n_1$ or (and) $n_2$ components. 
In most studies of this direction, the authors employ some potentials which are motivated by physics or mathematics, 
and they numerically solve the Euler-Lagrange equations or the Hamiltonian 
and get the desired solutions. 

In \cite{Garaud:2012pn}, the authors discussed a multi-band Ginzburg-Landau model. Especially, it forms 
several skyrmion exciations which exhibit molecular-like magnetic field configurations. The potential 
comprises of the standard Mexican hat and also the Josephson terms and the effects are thoroughly studied.  

The method examined in \cite{Piette:1992he} seems unique and independent from 
others. First the authors assume an existence of a static and exact $N-$centered 
solution. Next, they determine the form of the potential in order 
that the solution satisfies the Euler-Lagrange equation. They succeeded to get 
the analytical $N=2$ solution but failed to find $N>2$ ones
and then the problem was solved only numerically. 
Although any guiding principle to find the solution and the potential is absent 
in their discussion, the idea seems promising. 
In this paper, we try to construct the exact $N-$centered vortex solutions and 
the corresponding potentials. 
First, we introduce the $N-$centered ansatz which is essentially 
similar with one proposed in \cite{Piette:1992he}. Main difference is that our 
ansatz describes the time-dependent, travelling wave vortex solutions and these are exactly  solution of the corresponding submodel equation. 
Plugging them into the Euler-Lagrange equation of the model and 
we are able to construct the potential in order that the solution of the submodel becomes one of the equations itself. 
As a result, we can get the analytical vortex solution of the model which 
possesses an infinite number of the conserved quantities. The method is 
straightforwardly applicable to the other related soliton models such as 
the baby-Skyrme model, the Skyrme model, so on.

The paper is organized as follows.  In the next section we briefly describe the extended
Skyrme-Faddeev model. The equations of motion are also introduced in Sec. \ref{sec:vortex}.  
The method how to get the solutions of the integrable sector of the present model is discussed 
in Sec. \ref{sec:hamiltonian}.  
Sec.\ref{sec:integrable} is devoted for the zero-curvature conditions and the conservation of the currents. 
In Sec \ref{sec:numerics}, we show the
numerical solutions.  A brief summary is presented in Sec. \ref{sec:summary}.

\section{\label{sec:vortex}The model}
The Lagrangian density of the extended Skyrme Faddeev model reads (\ref{lagrangian}),
where unit vector $\vec{n}$, {\it i.e.} $\vec{n}\cdot\vec{n} = 1$, is a triplet of real scalar fields taking values on the sphere $S^2$.
The coupling constant $\mathcal{M}$ is dimensional whereas $e^2$ and $\beta$ are some dimensionless coupling constants.
The potential $V$ depends on all the components of the triplet $\vec{n}$. 
One can introduce the complex fields $u$ and $u^*$ using a stereographic projection(\ref{udef}) which leads to the following expression for the Lagrangian
\begin{align}
\mathcal{L} = 4\mathcal{M}^2 \frac{\partial_{\mu}u\partial^{\mu}u^*}{(1+|u|^2)^2}
+\frac{8}{e^2}\Big[\frac{(\partial_{\mu}u)^2(\partial_{\nu}u^*)^2}{(1+|u|^2)^4} \notag\\
+ (\beta e^2 -1)\frac{(\partial_{\mu}u\partial^{\mu}u^*)^2}{(1+|u|^2)^4}\Big] -V(u,u^*).
\label{lagrangian2}
\end{align}
If one set $\beta e^2 =1$, then the model possesses some lump shaped analytical solutions in absence of the potential $V$.
For the case of $\beta e^2 \neq 1$, a potential is needed to stabilize the solution~\cite{Ferreira:2011mz,Ferreira:2010zz}.
In this paper, we try to find an exact form of potential for some solutions with $\beta e^2 \neq 1$. \\
The Euler-Lagrange equations corresponding to (3) read 
\begin{equation}
(1+|u|^2)\partial^{\mu}\mathcal{K}_{\mu} - 2u^* \mathcal{K}_{\mu}\partial^{\mu}u
= - \frac{1}{4}(1+|u|^2)^3\frac{\partial V}{\partial u^*},
\label{sfeqn}
\end{equation}
together with the complex conjugated equation.
The symbol $\mathcal{K}_{\mu}$ stands for the expression
\begin{align}
\mathcal{K}_{\mu} \equiv \mathcal{M}^2 \partial_{\mu}u + \frac{4}{e^2}
\Bigl[\frac{(\partial_{\nu}u\partial^{\nu}u)\partial_{\mu}u^*}{(1+|u|^2)^2} \notag\\
 + \frac{(\beta e^2 -1)(\partial_{\nu}u\partial^{\nu}u^*)\partial_{\mu}u}{(1+|u|^2)^2} \Bigr].
\end{align}
We use the dimensionless polar coordinates ($t$, $\rho$, $\varphi$, $z$) defined as follows
\[
x_0 = ct,~x_1 = \rho \cos\varphi,~x_2 = \rho \sin\varphi,~x_3 = z.
\]
Here we choose $c=1$.
The metric is of the form
\begin{align}
(ds)^2 = \eta_{\mu\nu}dx^{\mu}dx^{\nu}= (dt)^2 - (d\rho)^2 - \rho^2 (d\varphi)^2 - (dz)^2,
\nonumber
\end{align}
where
\begin{equation*}
\eta_{\mu\nu} = \begin{pmatrix}
1 & 0 & 0 & 0  \\
0 & -1 & 0 & 0  \\
0 & 0 & -\rho^2 & 0  \\
0 & 0 & 0 & -1  \\
\end{pmatrix},
~~~~\det[\eta] = -\rho^2.
\end{equation*}
All classical configurations are characterised by the following topological charge
\begin{align}
Q&=\frac{1}{4\pi}\int\vec{n}\cdot(\partial_{\mu}\vec{n}\times\partial_{\nu}\vec{n})d^2x  \nonumber \\
&=\frac{i}{2\pi}\int\frac{(\partial_{\rho}u\partial_{\varphi}u^*-\partial_{\varphi}u\partial_{\rho}u^*)}{(1+|u|^2)^2}\rho d\rho d\varphi.
\label{charge}
\end{align}
This integer $Q$ is associated to the vortex, and defined as the winding number of the map from any circle on the $x$-$y$ plane
 centered at the $z$-axis.

\section{\label{sec:hamiltonian}Construction of the exact solutions}
\subsection{The general formula}
We would like to examine solutions which satisfy so called zero curvature condition~\cite{Alvarez:1997ma}.
The zero curvature condition of the model reads
\begin{align}
(\partial_{\mu}u)^2=0.
\label{zccondition}
\end{align}
As we shall describe at Sec.\ref{sec:integrable} in detail, 
the solution possesses an infinite set of conserved currents.
Within the condition (\ref{zccondition}), the equation of motion (\ref{sfeqn}) 
can be written as
\begin{align}
\frac{\partial V}{\partial u^*}=
\frac{1}{(1+|u|^2)^2}\partial^{\mu}\biggl[\frac{16(\beta e^2-1)
(\partial_{\nu}u\partial^{\nu}u^*)
\partial_{\mu}u}{e^2(1+|u|^2)^2}\biggr].
\label{difeq_pot}
\end{align}
In order to find the explicit form of $V$, it is not straightforward, or not possible to solve the partial differential equation (\ref{difeq_pot}).
In the following, we propose an ansatz to find a good candidate of $V$ which consists of a few steps: 
\begin{itemize}
\item[(i)]~we construct a $N$-centered solution $u_N$ which satisfies the zero curvature condition (\ref{zccondition}),
\item[(ii)]~substituting the solution found in (i) into (\ref{difeq_pot}), we get the derivative of a potential
\begin{eqnarray}
&&\frac{\partial \bar{V}}{\partial u_N^*} \nonumber \\
&&=
\frac{1}{(1+|u_N|^2)^2}\partial^{\mu}\biggl[\frac{16(\beta e^2-1)
(\partial_{\nu}u_N\partial^{\nu}u_N^*)
\partial_{\mu}u_N}{e^2(1+|u_N|^2)^2}\biggr],
\nonumber 
\end{eqnarray}
\item[(iii)]~we introduce a candidate of $V_N$ and examine its derivative $\dfrac{\partial V_N}{\partial u^*}$,
\item[(iv)]~we confirm that $V_N$ is correct by comparing $\dfrac{\partial V_N}{\partial u^*}$ and $\dfrac{\partial \bar{V}}{\partial u^*}$.
\end{itemize}

As a simple example, we consider a solution of the form
\begin{align}
u_N=\frac{\rho e^{i\varphi}-\xi_1}{a}\frac{\rho e^{i\varphi}-\xi_2}{a}\frac{\rho e^{i\varphi}-\xi_3}{a} \cdots \frac{\rho e^{i\varphi}-\xi_N}{a}
\label{solution_ex}
\end{align}
where $\xi_1,\xi_2,\cdots, \xi_N (\in \mathbb{C})$ have a meaning of position of the center of each constituent, and $a$ is a scale parameter.
For simplicity, we set the centers $\xi_k\equiv ce^{i(\alpha+\frac{2\pi(k-1)}{N})}$,where $k=1,\cdots,N$,
and $\alpha$ and $c$ are arbitrary constants then the solution
(\ref{solution_ex}) is contracted in the form
\begin{align}
u_N=\frac{1}{a^N}\left(\rho^N e^{iN\varphi}-\xi_1^N\right).
\label{somesolution}
\end{align}
Substituting (\ref{somesolution}) into (\ref{difeq_pot}), we get
\begin{align}
&\frac{\partial \bar{V}}{\partial u_N^*}=-\frac{64 N^3 (\beta  e^2-1)
\Delta^{*1-\frac{2}{N}} \Delta^{2-\frac{2}{N}}}{a^4e^2 \left(1+|u_N|^2\right)^5}\notag \times \\
&\biggl[1+|u_N|^2-N\biggl\{1-2\Bigl(\frac{\xi^*_1}{a}\Bigr)^{N}\biggl(\Delta-\Bigl(\frac{\xi_1}{a}\Bigr)^N\biggr)-|u_N|^2\biggr\}\biggr],
\label{pot_coordinate}
\end{align}
where $\Delta \equiv \Bigl(\dfrac{\rho}{a}\Bigr)^N e^{iN\varphi}$.
We rewrite (\ref{pot_coordinate}) by using 
\begin{align}
\Delta \to u_N+\left(\frac{\xi_1}{a}\right)^N,~~
\Delta^* \to u_N^*+\left(\frac{\xi^*_1}{a}\right)^N
\end{align}
and then
\begin{align}
&\dfrac{\partial \bar{V}}{\partial u_N^*}= 
\frac{-2\lambda}{N (1+|u_N|^2)^5} \notag\\
&\times \biggl[\biggl(u_N+\Bigl(\frac{\xi_1}{a}\Bigr)^N\biggr)^{2-\frac{2}{N}}
\biggl(u_N^*+\Bigl(\frac{\xi_1^*}{a}\Bigr)^N\biggr)^{1-\frac{2}{N}}\notag\\
&\times\biggl\{1+|u_N|^2-N\biggl(1-2\Bigl(\frac{\xi_1^*}{a}\Bigr)^N u_N-|u_N|^2\biggr)\biggr\} \biggr]
\label{dd_pot}
\end{align}
with
\begin{align}
\lambda=\frac{-32N^4(\beta e^2-1)}{a^4 e^2}.
\end{align}
Here a candidate of $V_N$ is introduced as
\begin{align}
V_N= \frac{\lambda}{(1+|u_N|^2)^4}\biggl(u_N+\Bigl(\frac{\xi_1}{a}\Bigr)^N\biggr)^{a}\biggl(u_N^*+\Bigl(\frac{\xi_1^*}{a}\Bigr)^N\biggr)^{b},
\end{align}
where $\lambda, a$ and $b$ are arbitrary constants. 
Comparing $\dfrac{\partial V_N}{\partial u^*}$ and $\dfrac{\partial \bar{V}}{\partial u^*}$, 
we can fix the constants and obtain the form of the potential
\begin{align}
V_N= \frac{\lambda}{(1+|u_N|^2)^4}\biggl(u_N+\Bigl(\frac{\xi_1}{a}\Bigr)^N\biggr)^{2-\frac{2}{N}}
\biggl(u_N^*+\Bigl(\frac{\xi^*_1}{a}\Bigr)^N\biggr)^{2-\frac{2}{N}}.
\label{potential_ex}
\end{align}  
In the next subsection, we consider the solutions with the several winding numbers (the centers) $N$ 
and the corresponding potentials.

\begin{figure}[t]
\includegraphics[width=7cm]{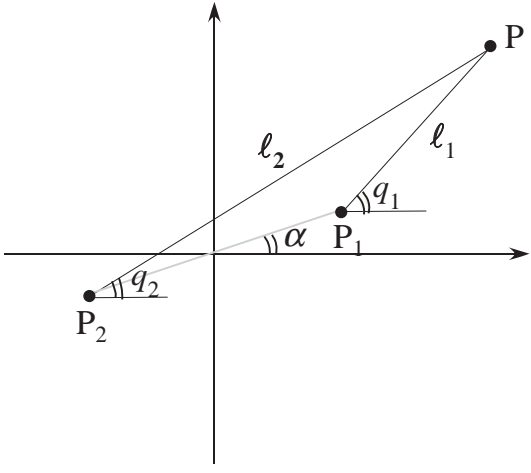}%
\caption{Schematic picture for the two-centered vortex solution. The each vortex sits on 
$(c,\alpha)$ and $(c,\alpha+\pi)$. 
Independently, they have the own radius and the azimuthal angle $(\ell_i,q_i),~i=1,2$, which 
is used to describe the structure of each vortex. }
\label{fig:two-centered}
\end{figure}

\subsection{Examples of the solutions and the corresponding potentials}
For the construction of the solution and the potential, 
there have been already some studies based on the scheme~\cite{Ferreira:2011mz,Piette:1992he}.
First we present the results and then focus on several new multi-centered holomorphic solutions and 
the corresponding potentials.

\subsubsection{The lump shaped solution and the corresponding potential}
First we examine the case of the lump solution and see how the scheme works.
The holomorphic vortex solution with the winding number $Q=1$ is \cite{vortexlaf}
\begin{align}
u_1=\Big(\frac{\rho}{a}\Big) e^{i[\varphi+k(t+z)]},
\label{sol_1center}
\end{align}
where $a$ is an arbitrary scale parameter.
Note that it satisfies the zero curvature condition (\ref{zccondition})
and possesses an infinite set of the conserved quantities.
It is easy to show that the following formulas hold
\begin{align}
\partial_\mu u_1\partial^\mu u_1^*=-\frac{2}{\rho^2}|u_1|^2,~~\partial^\mu \partial_\mu u_1=0.
\end{align}
Substituting (\ref{sol_1center}) into (\ref{difeq_pot}), we get
\begin{align}
\frac{\partial \bar{V}}{\partial u^*}\biggl|_{u=u_1}=-4\lambda\frac{u_1}{(1+|u_1|^2)^5}.
\label{pot_1center0}
\end{align}
On the other hand, plugging (\ref{sol_1center}) into (\ref{potential_ex}), we obtain
\begin{align}
V_1=\lambda\frac{1}{(1+|u_1|^2)^4}
\label{potential_1}
\end{align}
and by taking derivative of this, we obtain
\begin{align}
\frac{\partial V_1}{\partial u^*}\biggl|_{u=u_1}=-4\lambda\frac{u_1}{(1+|u_1|^2)^5}.
\label{pot_1center}
\end{align}
(\ref{pot_1center0}) and $(\ref{pot_1center})$ are apparently equal and then (\ref{potential_1})
is valid in this case. In fact, it is the same as the one which was found in \cite{Ferreira:2011mz}.

\subsubsection{The two-centered solution and the potential}
The solution with two-center can be written as a straightforward 
generalization of (\ref{sol_1center}) such as 
\begin{align}
u_2&=\Bigl(\frac{\ell_1}{a}\Bigr) \Bigl(\frac{\ell_2}{a}\Bigr) 
e^{i[(q_1+q_2)+k(t+z)]}\,.
\label{sol_2center}
\end{align}
The solution possesses two ``radii $\ell_i$'' and ``azimuthal angles $q_i$'' with a scale parameter $a$,
such that $\ell_i$ are a distances from a reference point $P$ and $q_i$ are angles measured from the 
horizontal axis. 
If we assume that each center is located at $(c,\alpha)$ and $(c,\alpha+\pi)$ in polar coordinate (Fig.\ref{fig:two-centered}), 
 $(\ell_i,q_i)$ are written in terms of the coordinate of $P$: $(\rho,\varphi)$ and $(c,\alpha )$ as 
\begin{align}
\ell_1 &:=\sqrt{\rho^2-2\rho c \cos(\varphi-\alpha)+c^2},\notag \\
q_1 &:=\arctan\biggr[\frac{\rho\sin\varphi-c \sin\alpha}{\rho \cos\varphi-c\cos\alpha}\biggl]; \notag \\
\ell_2 &:=\sqrt{\rho^2+2\rho c \cos(\varphi-\alpha)+c^2}, \notag \\
q_2 &:=\arctan\biggr[\frac{\rho\sin\varphi+c \sin\alpha}{\rho \cos\varphi+c\cos\alpha}\biggl].
\label{coordinatesrelation2}
\end{align}
After a bit lengthy calculation, we find form of the solution 
\begin{equation}
u_2=\tilde{\rho}^{\hspace{3pt}2} e^{i[2\varphi+k(t+z)]}
-\tilde{c}^{\hspace{3pt}2} e^{i[2\alpha+k(t+z)]}, 
\label{sol_2center}
\end{equation}
where $\tilde{\rho}=\rho/a, \tilde{c}=c/a$. In the following, we omit `~$\tilde{~}$~' for simplicity. 
The static solution of (\ref{sol_2center}),i.e., the case $k=0$, was already found as a static planar solution of 
the baby-Skyrme model \cite{Piette:1992he}, in which the authors mainly focused on the effective force 
between the two constituents by using the solution.  
Of course (\ref{sol_2center}) is also a solution of our model. It is easy to show that the following formulas hold
\begin{align}
\partial_\mu u_2\partial^\mu u_2^*=-\frac{2}{\rho^2}|u_2|^2,~~\partial^\mu \partial_\mu u_2=0.
\end{align}
Substituting (\ref{sol_2center}) into (\ref{difeq_pot}), we get
\begin{align}
\frac{\partial \bar{V}}{\partial u^*}\biggl|_{u=u_2}=-4\lambda\frac{(u_2+c^2)(-1+3|u_2|^2+4c^2u_2)}{(1+|u_2|^2)^5}.
\label{pot_u2}
\end{align}
On the other hand, substituting (\ref{sol_2center}) into (\ref{potential_ex}), we obtain
\begin{align}
V_2=\lambda\frac{(u_2+c^2)(u_2^*+c^2)}{(1+|u_2|^2)^4}
\label{potential_2u}
\end{align}
and the derivative
\begin{align}
\frac{\partial V_2}{\partial u^*}\biggl|_{u=u_2}=-4\lambda\frac{(u_2+c^2)(-1+3|u_2|^2+4c^2u_2)}{(1+|u_2|^2)^5}.
\label{pot_2centeru}
\end{align}
Since (\ref{pot_u2}) and (\ref{pot_2centeru}) coincide then (\ref{potential_2u})
is valid. In terms of the $\vec{n}$ field, the potential has the form 
\begin{align}
&V_2=\frac{\lambda}{16}\Big\{n_1+in_2
+c^{2} e^{i[2\alpha+k(t+z)]}(1-n_3)\Big\} \nonumber \\ 
&\times\Big\{n_1-in_2+c^{2} e^{-i[2\alpha+k(t+z)]}(1-n_3)\Big\}(1-n_3)^2
\label{pot_2center}
\end{align}
which coincides with the static case $k=0$ in \cite{Piette:1992he}.

\begin{figure}[t]
\includegraphics[width=7cm]{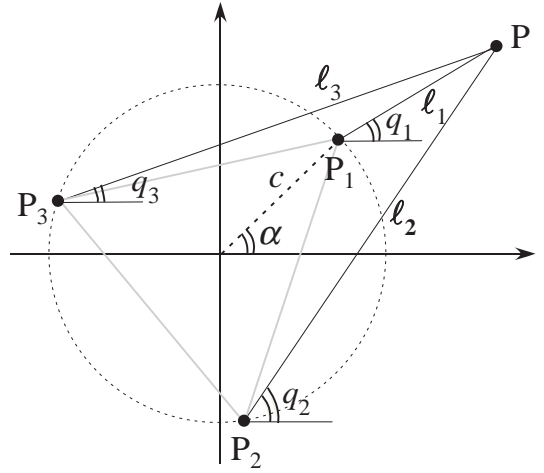}%
\caption{Schematic picture for the three-centered vortex solution. The each vortex sits on 
$(c,\alpha), (c,\alpha+\frac{2}{3}\pi), (c,\alpha+\frac{4}{3}\pi)$, respectively. 
Almost independently, they have the own radius and the azimuthal angle $(\ell_i,q_i),~i=1,2,3$, which 
is used to describe the structure of the vortex. }
\label{fig:three-centered}
\end{figure}

\subsubsection{The three-centered solution}
We would like to apply the scheme for more complex systems.  
The first non-trivial case is with the three-center.
Construction of a holomorphic solution with three isolated centers is not 
straightforward and a geometrical understanding is helpful.  
In Fig.\ref{fig:three-centered}, we schematically draw a  picture about 
the location of the core of vortices and the coordinates. 
As before we define a reference point $P$ which has a component $(\rho,\varphi)$ of a two dimensional 
polar coordinate. 
We consider the case that the vortices sit on the three tops of an equilateral triangle. 
The tops of the triangle $P_i,i=1,2,3$ are located on a circle with 
the radius $c$. If $P_1$ has a component $(c,\alpha)$ in the polar coordinate, 
the remaining $P_2,P_3$ are then 
$(c,\alpha+\frac{2}{3}\pi), (c,\alpha+\frac{4}{3}\pi)$, respectively. In order to construct the 
solution, according to the case of $n=1$ (\ref{sol_1center})
we introduce ``the radius $\ell_i$'' and ``the azimuthal angle $q_i$'' of $P_i$, such that $\ell_i$ are distances 
between $P$ and $P_i$. $q_i$ are angles measured from the horizontal axis. 
The solution with three-center is described in terms of these coordinates such as
\begin{align}
u_3&=\Bigl(\frac{\ell_1}{a}\Bigr) \Bigl(\frac{\ell_2}{a}\Bigr) \Bigl(\frac{\ell_3}{a}\Bigr) 
e^{i[(q_1+q_2+q_3)+k(t+z)]}.
\end{align}
The relation between the $(\ell_i,q_i)$  and the polar coordinate $(\rho,\varphi)$,  $(c,\alpha)$  are
\begin{align}
\ell_1 &:=\sqrt{\rho^2-2\rho c \cos(\varphi-\alpha)+c^2},\notag \\
q_1 &:=\arctan\biggr[\frac{\rho\sin\varphi-c \sin\alpha}{\rho \cos\varphi-c\cos\alpha}\biggl]; \notag \\
\ell_2 &:=\sqrt{\rho^2+\rho c\cos(\varphi-\alpha)-\sqrt{3}\rho c \sin(\varphi-\alpha)+c^2}, \notag \\
q_2 &:=\arctan\biggr[\frac{2\rho \sin\varphi-\sqrt{3}c\cos\alpha+c\sin\alpha}{2\rho\cos\varphi+c\cos\alpha+\sqrt{3}c\sin\alpha}\biggl]; \notag \\
\ell_3 &:=\sqrt{\rho^2+\rho c\cos(\varphi-\alpha)+\sqrt{3}\rho c \sin(\varphi-\alpha)+c^2},\notag  \\
q_3 &:=\arctan\biggr[\frac{2\rho \sin\varphi+\sqrt{3}c\cos\alpha+c\sin\alpha}{2\rho\cos\varphi+c\cos\alpha-\sqrt{3}c\sin\alpha}\biggl].
\label{coordinatesrelation}
\end{align}
Finally, we write down the form of the solution as
\begin{align}
u_3=\rho^{3}e^{i[3\varphi+k(t+z)]}
-c^{3}e^{i[3\alpha+k(t+z)]}.
\label{sol_3center}
\end{align}
Note that the solution (\ref{sol_3center}) is holomorphic and then, it satisfies the zero curvature condition (\ref{zccondition})
\begin{align}
(\partial_{\mu}u_3)^2 &= \eta^{\mu\nu}\partial_{\mu}u_3\partial_{\nu}u_3 \notag \\
&=(\partial_{t}u_3)^2-(\partial_{\rho}u_3)^2
-\frac{1}{\rho^2}(\partial_{\varphi}u_3)^2-(\partial_{z}u_3)^2 \notag \\
&=(iku_3)^2-\Bigl(\frac{3}{\rho}\psi\Bigr)^2-\frac{1}{\rho^2}(3i\psi)^2-(iku_3)^2 \notag\\
&=0,
\end{align}
where $\psi\equiv\rho^{\hspace{3pt}3}e^{3i\varphi}e^{ik(t+z)}$.
Substituting (\ref{sol_3center}) into (\ref{difeq_pot}), we find the potential
\begin{align}
V_3&=\frac{\lambda}{16}\Big\{n_1+in_2+c^{3} 
e^{i[3\alpha+k(t+z)]}(1-n_3)\Big\}^{\frac{4}{3}}\notag \\
&\times \Big\{n_1-in_2+c^{3} 
e^{-i[3\alpha+k(t+z)]}(1-n_3)\Big\}^{\frac{4}{3}}(1-n_3)^{\frac{4}{3}}.
\label{pot_3center}
\end{align}

\subsubsection{The $N$-centered solution}
The generalization is now almost straightforward. We assume the center of the vortices sit on the 
$N$-th tops of a equilateral polygon and 
the $N$-centered solution should be written as
\begin{equation}
u_N=\rho^{N} 
e^{i[N\varphi+k(t+z)]}-c^{N} e^{i[N\alpha+k(t+z)]}.
\label{sol_ncenter}
\end{equation}
The solution (\ref{sol_ncenter}) still satisfies the zero curvature condition 
(\ref{zccondition}) and then has an infinite number of conserved quantities. 
We get the following form of the potential for $N$-centered solution
\begin{align}
&V_N=\frac{\lambda}{16}\Big\{n_1+in_2+c^{N} 
e^{i[N\alpha+k(t+z)]}(1-n_3)\Big\}^{2-\frac{2}{N}} \notag \\ 
&\hspace{1cm}\times \Big\{n_1-in_2+c^{N} e^{-i[N\alpha+k(t+z)]}
(1-n_3)\Big\}^{2-\frac{2}{N}} \notag \\
&\hspace{5cm}\times (1-n_3)^{\frac{4}{N}}.
\label{pot_ncenter}
\end{align}
For the special case such as $c=0$ the potential (\ref{pot_ncenter}) becomes $V_N\to (1+n_3)^{2-2/N}(1-n_3)^{2+2/N}$, 
which perfectly agree with the result that we previously found in \cite{Ferreira:2011mz}.

\begin{figure}[t]
\includegraphics[width=7cm]{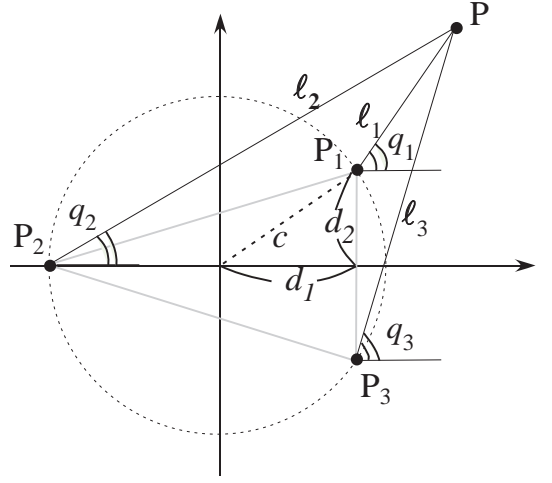}%
\caption{Schematic picture for the isosceles triangle vortex solution.
Independently, they have the own radius and the azimuthal angle $(\ell_i,q_i),~i=1,2,3$, which 
is used to describe the structure of the vortex. }
\label{isoscelestriangle}
\end{figure}

\begin{figure*}[t]
  \includegraphics[width=6cm]{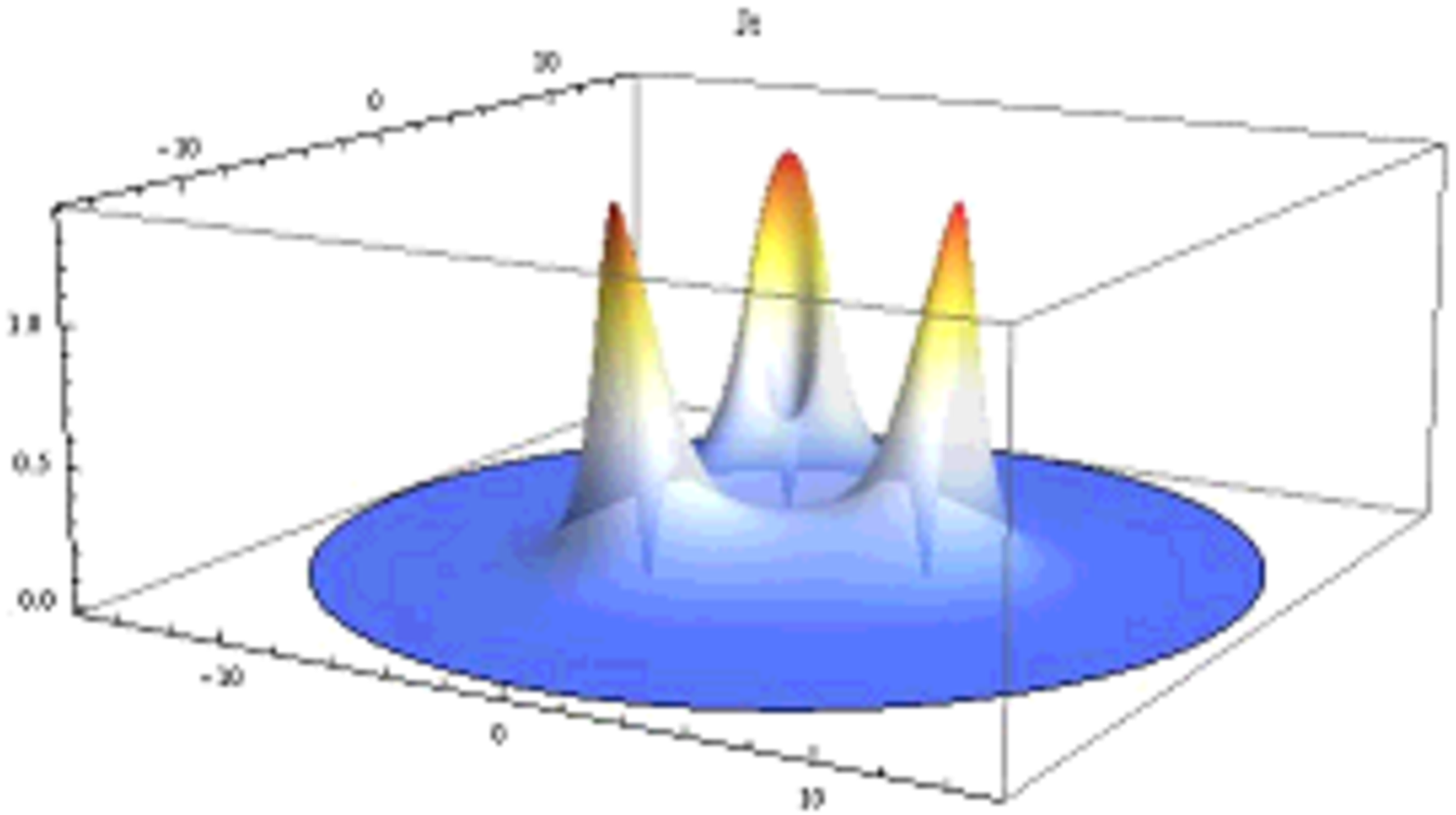}
\includegraphics[width=6cm]{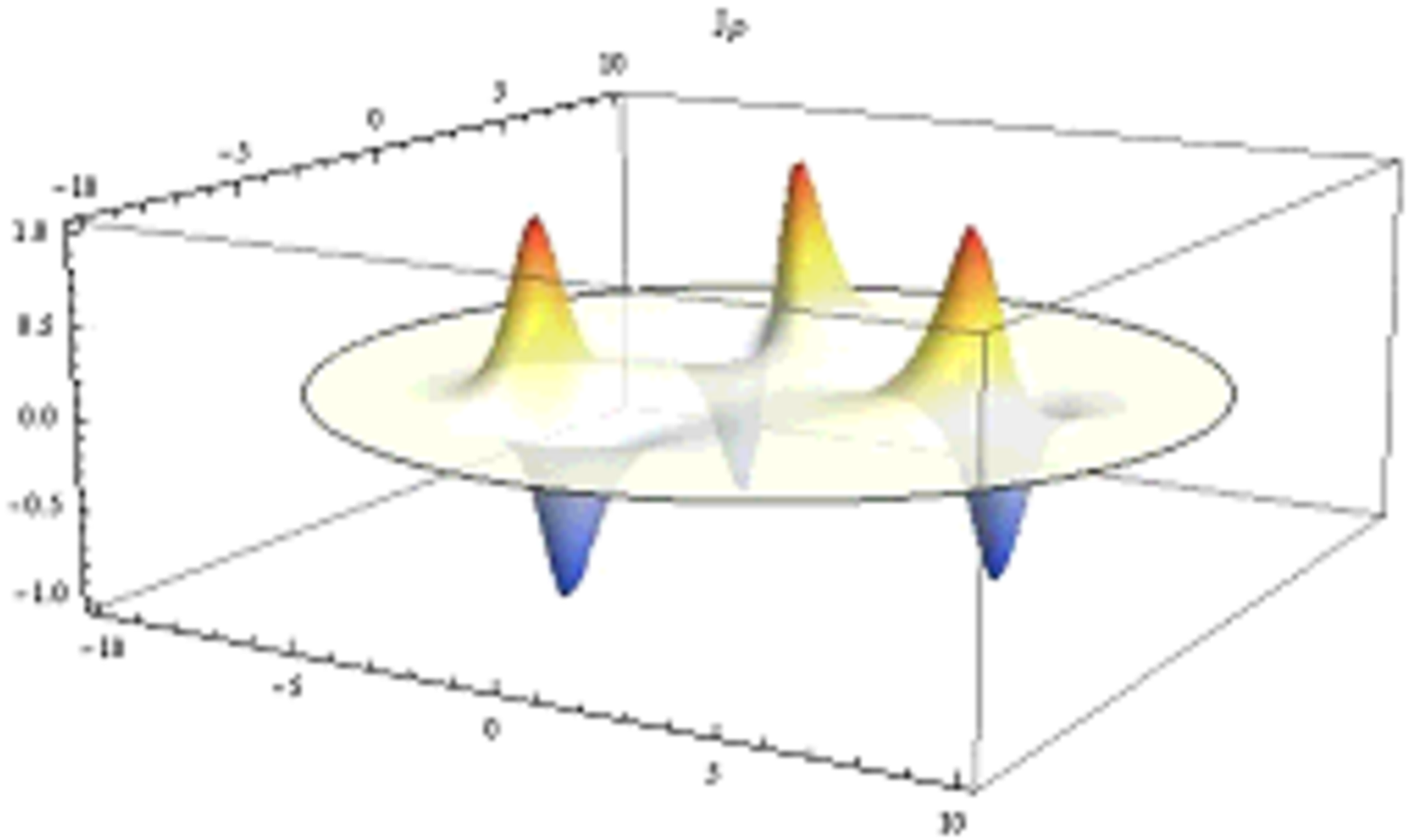}\\
 \includegraphics[width=6cm]{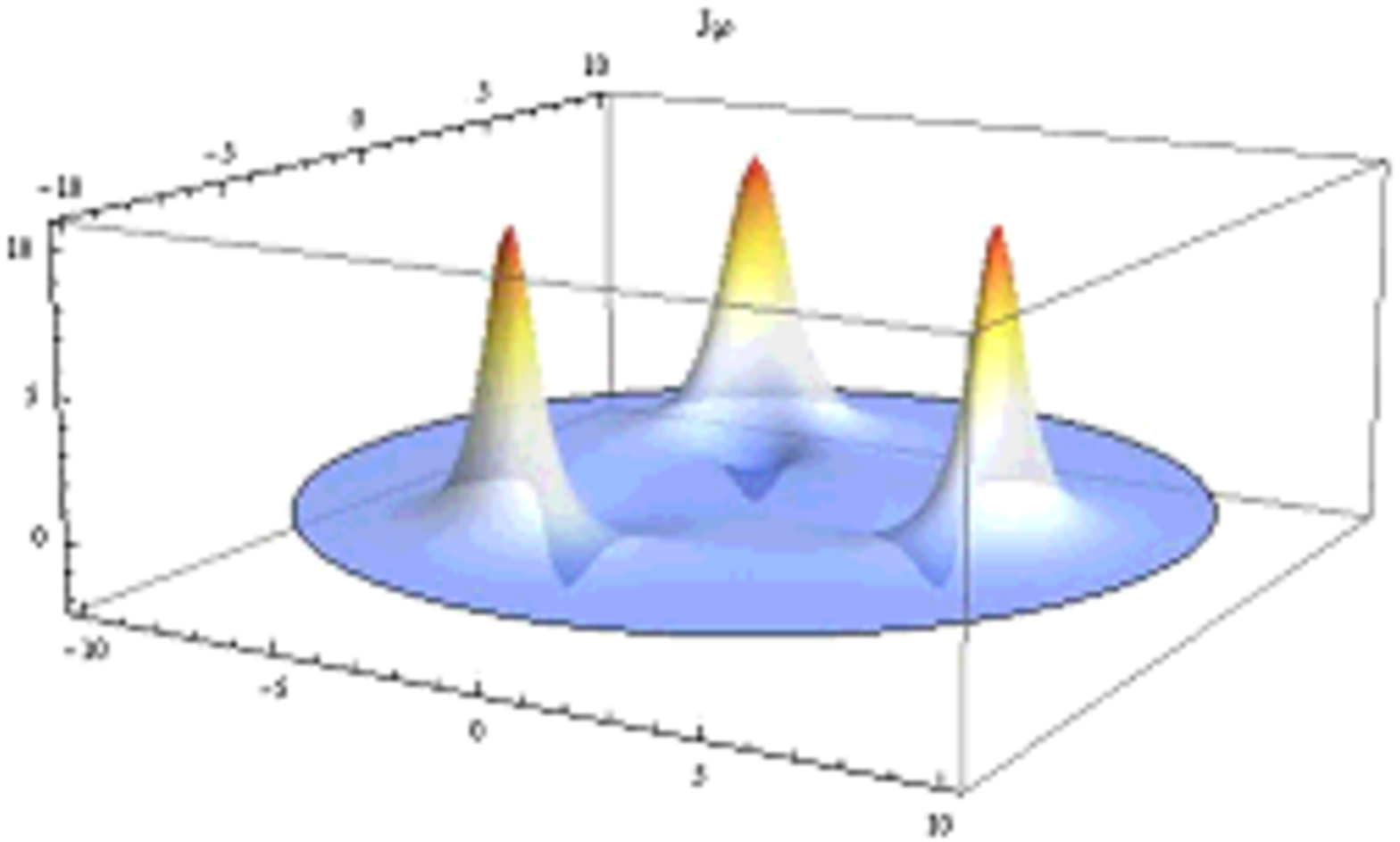}
\includegraphics[width=6cm]{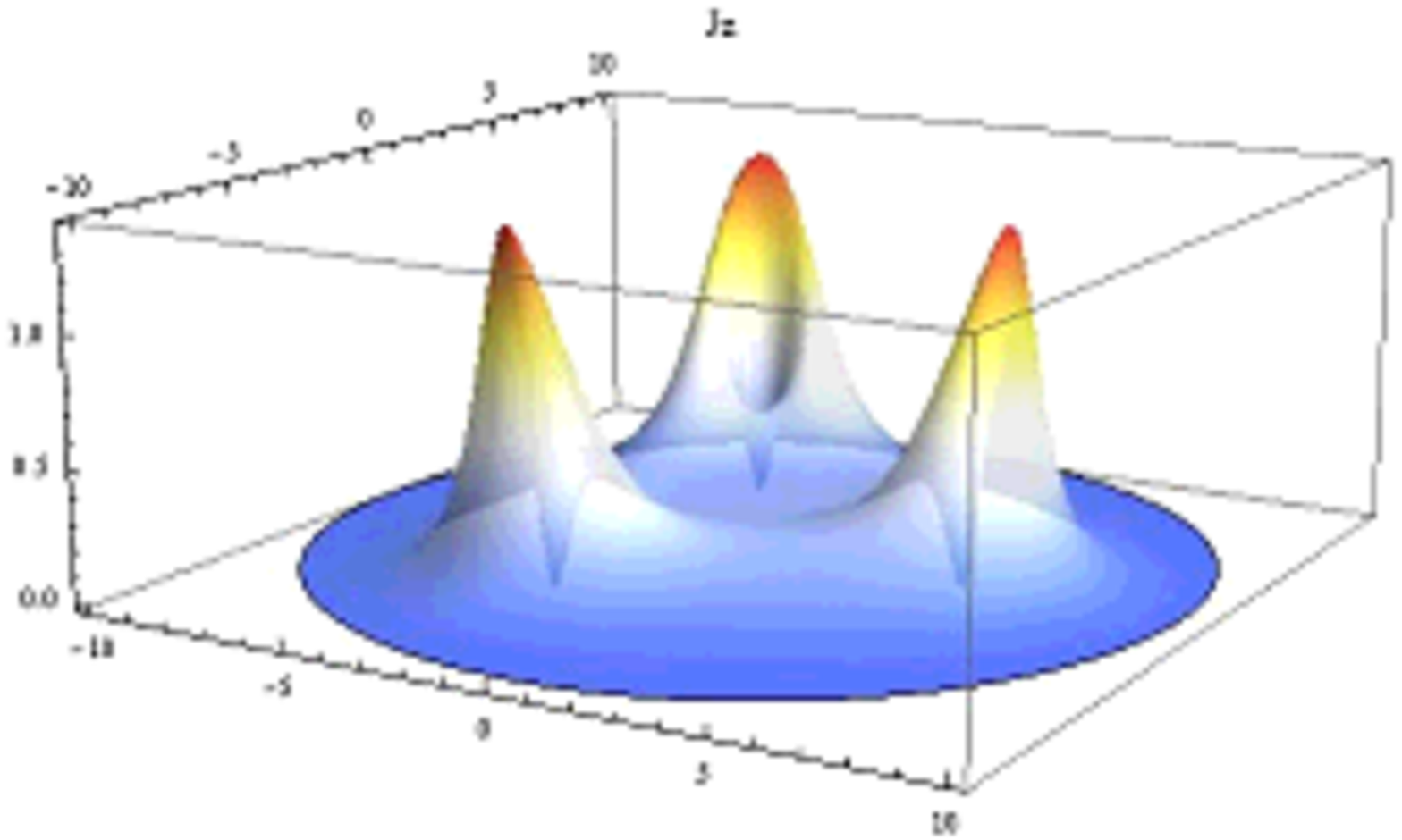}
 \caption{An U(1) current with the components $J_{t}$ (top left), $J_{\rho}$ (top right), 
$J_{\varphi}$ (bottom left) and $J_{z}$ (bottom right) (in unit of $-1/8{\cal M}^2$) with $N=3$ for  
$\beta=-2.0, e^2=-1.0, k=0.0$ and $c=1.0$.}
\label{currentdensity}
\end{figure*}

\begin{figure*}[htbp]
  \includegraphics[width=6cm]{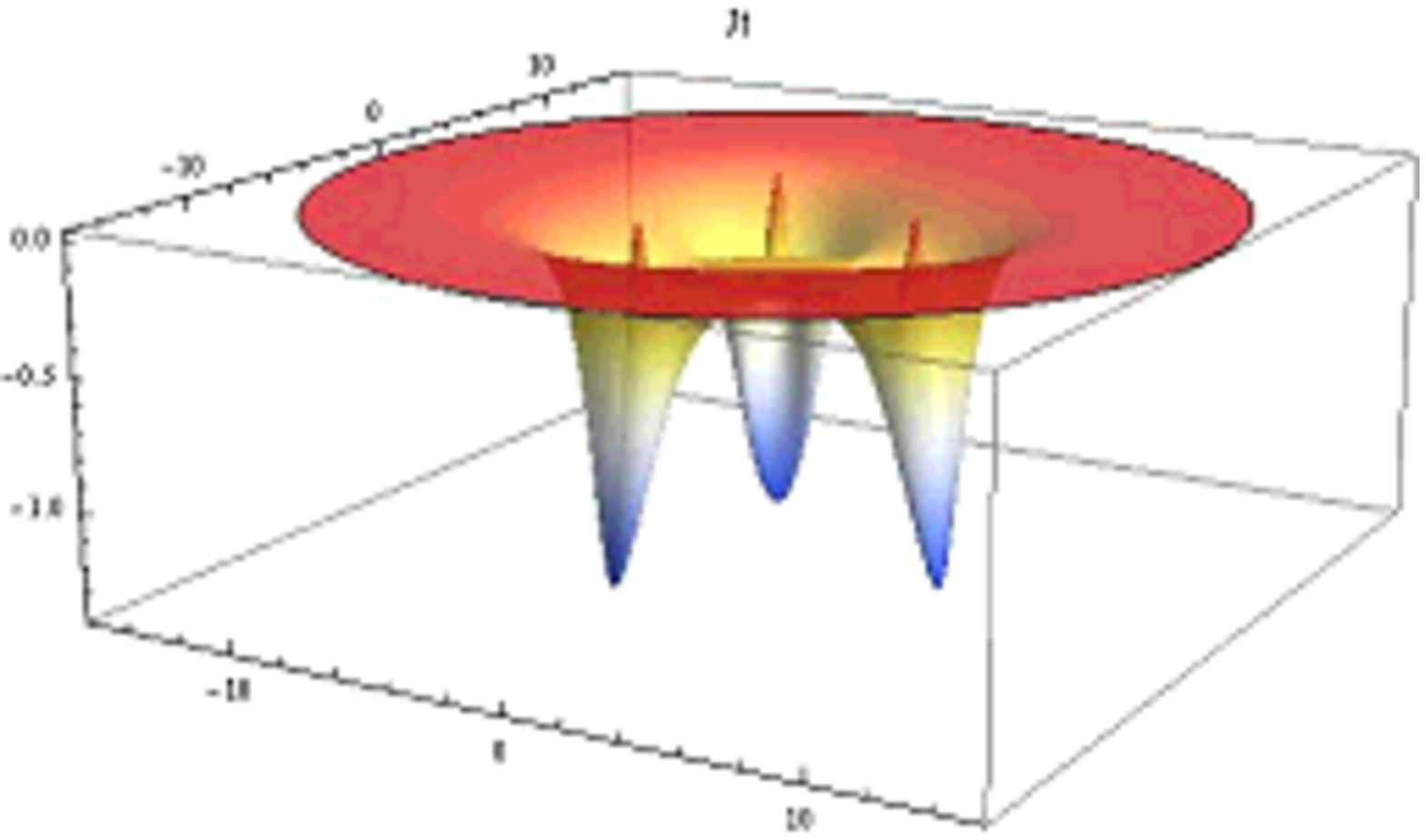}
\includegraphics[width=6cm]{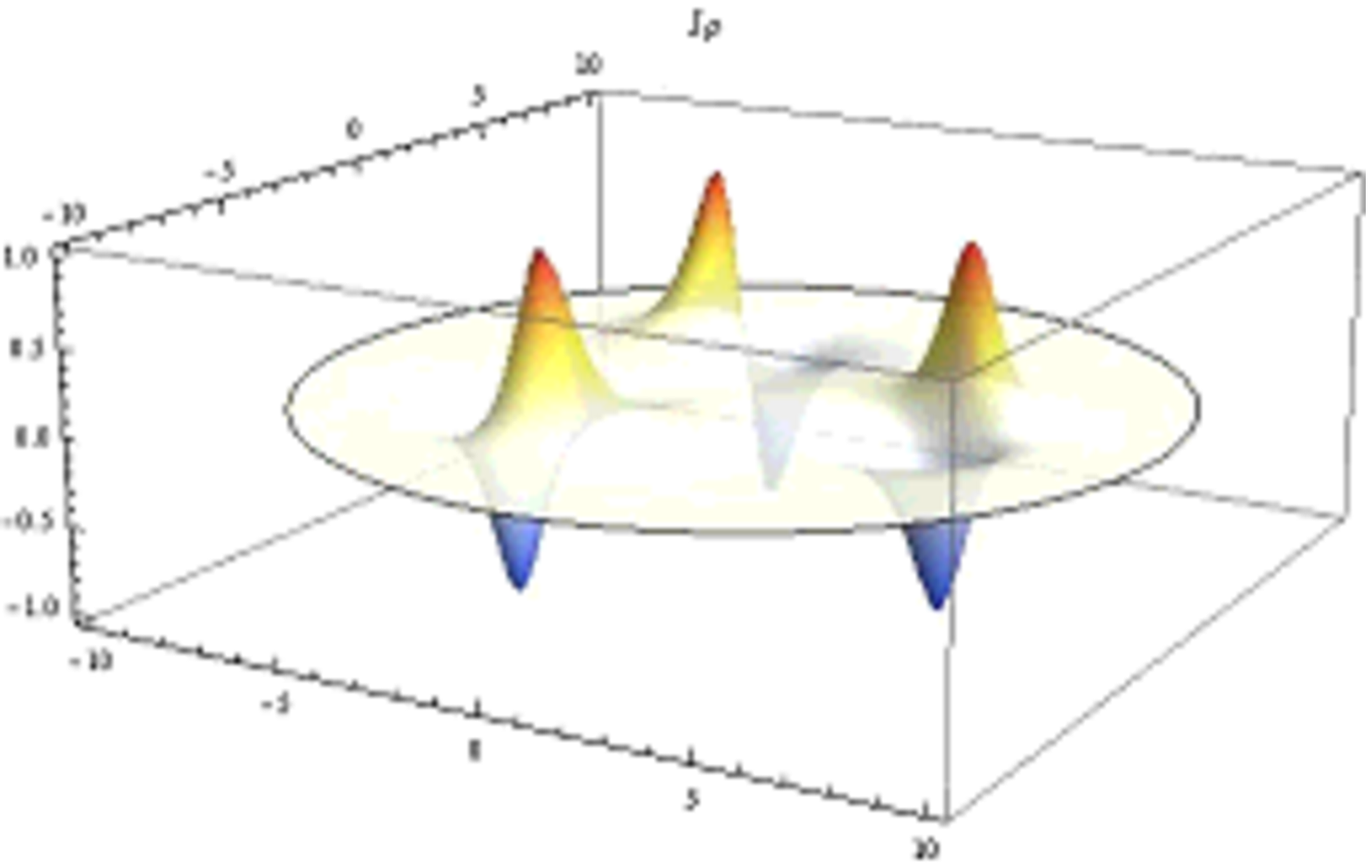}\\
 \includegraphics[width=6cm]{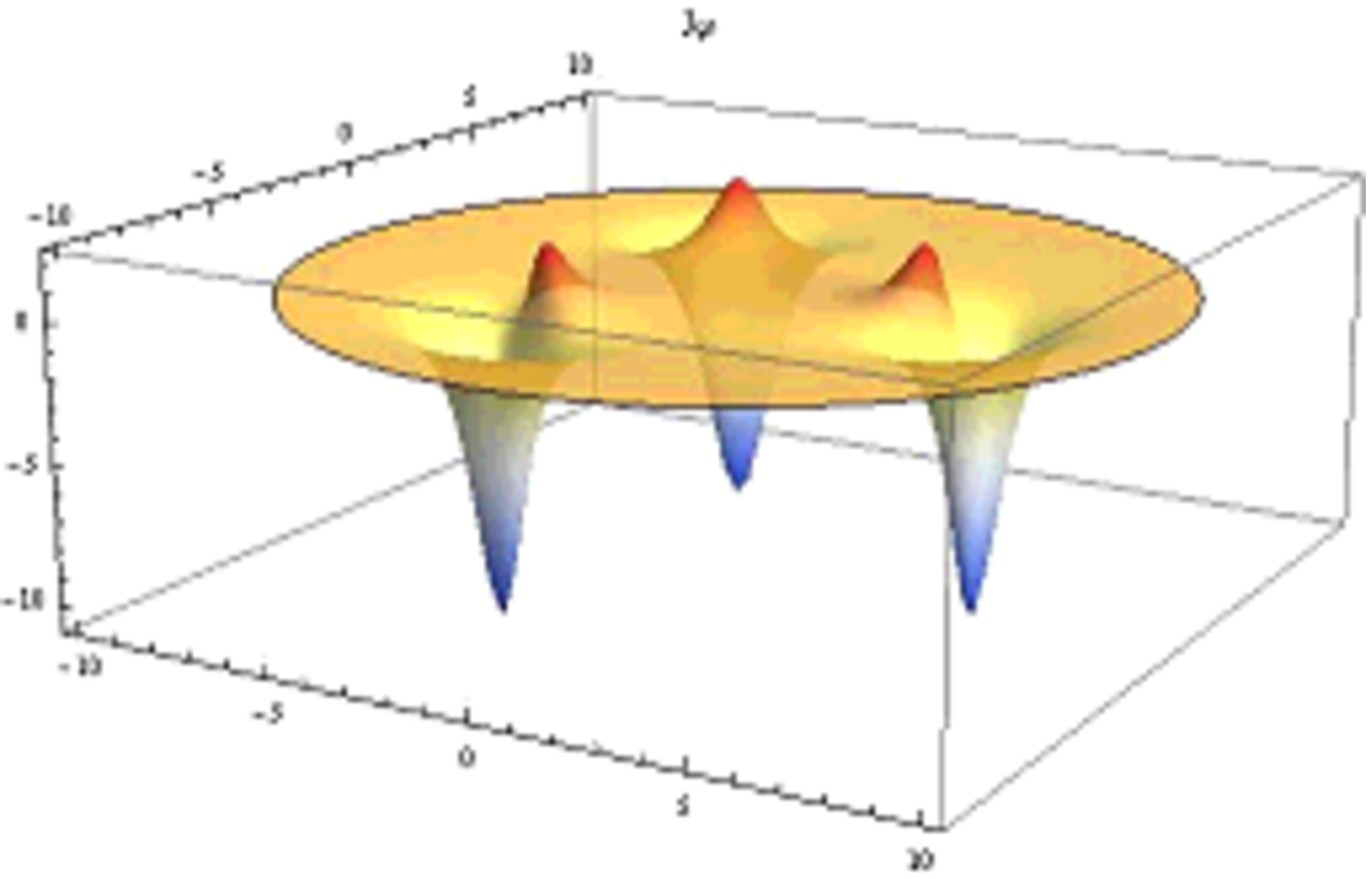}
\includegraphics[width=6cm]{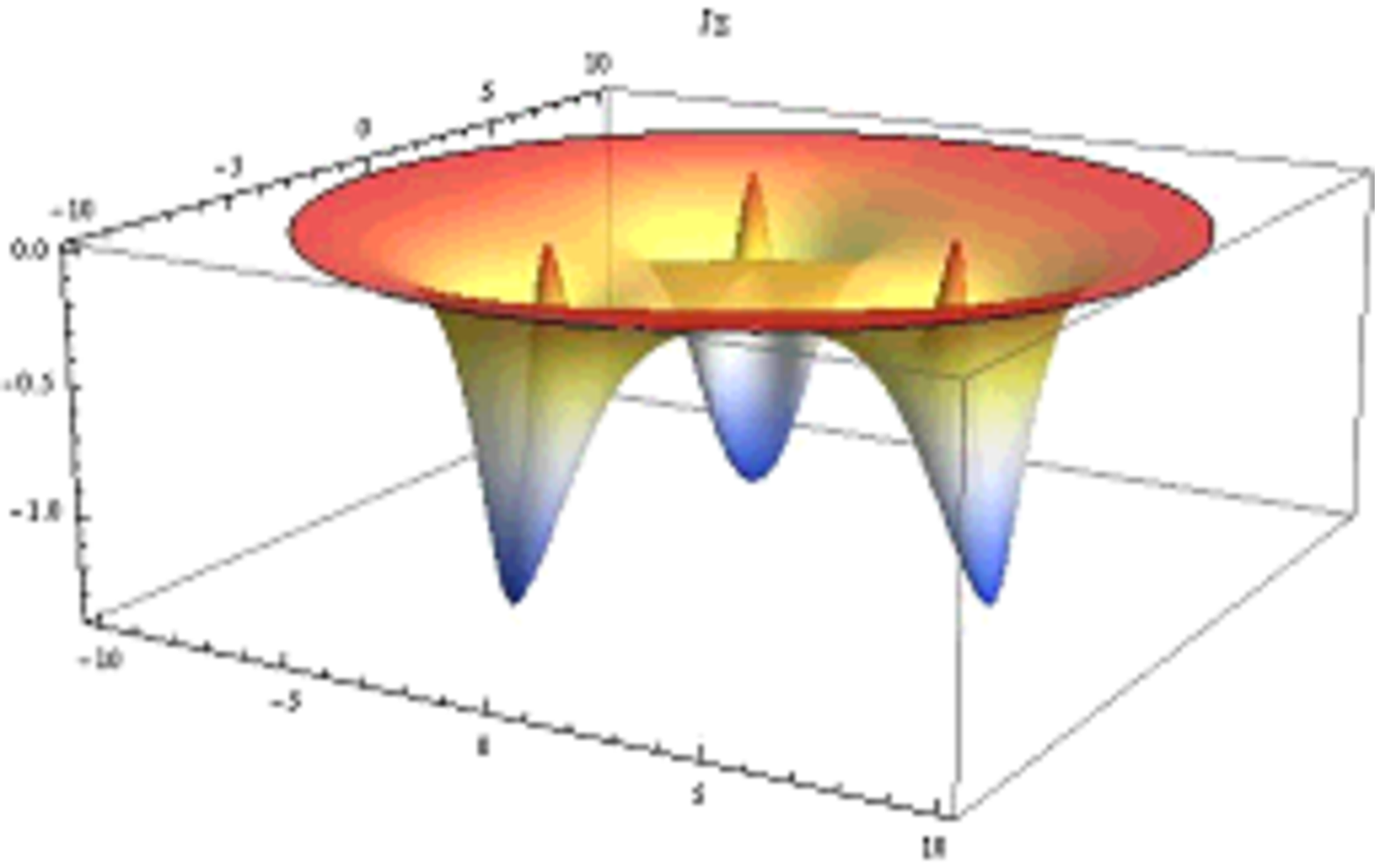}
 \caption{The components of a current which is defined by (\ref{currentg2}): $J_{t}$ (top left), $J_{\rho}$ (top right), 
$J_{\varphi}$ (bottom left) and $J_{z}$ (bottom right) (in unit of $-1/8{\cal M}^2$) with $N=3$ for  
$\beta=-2.0, e^2=-1.0, k=0.0$ and $c=1.0$.}
\label{currentdensity2}
\end{figure*}

\begin{figure}[h]
\includegraphics[width=5cm]{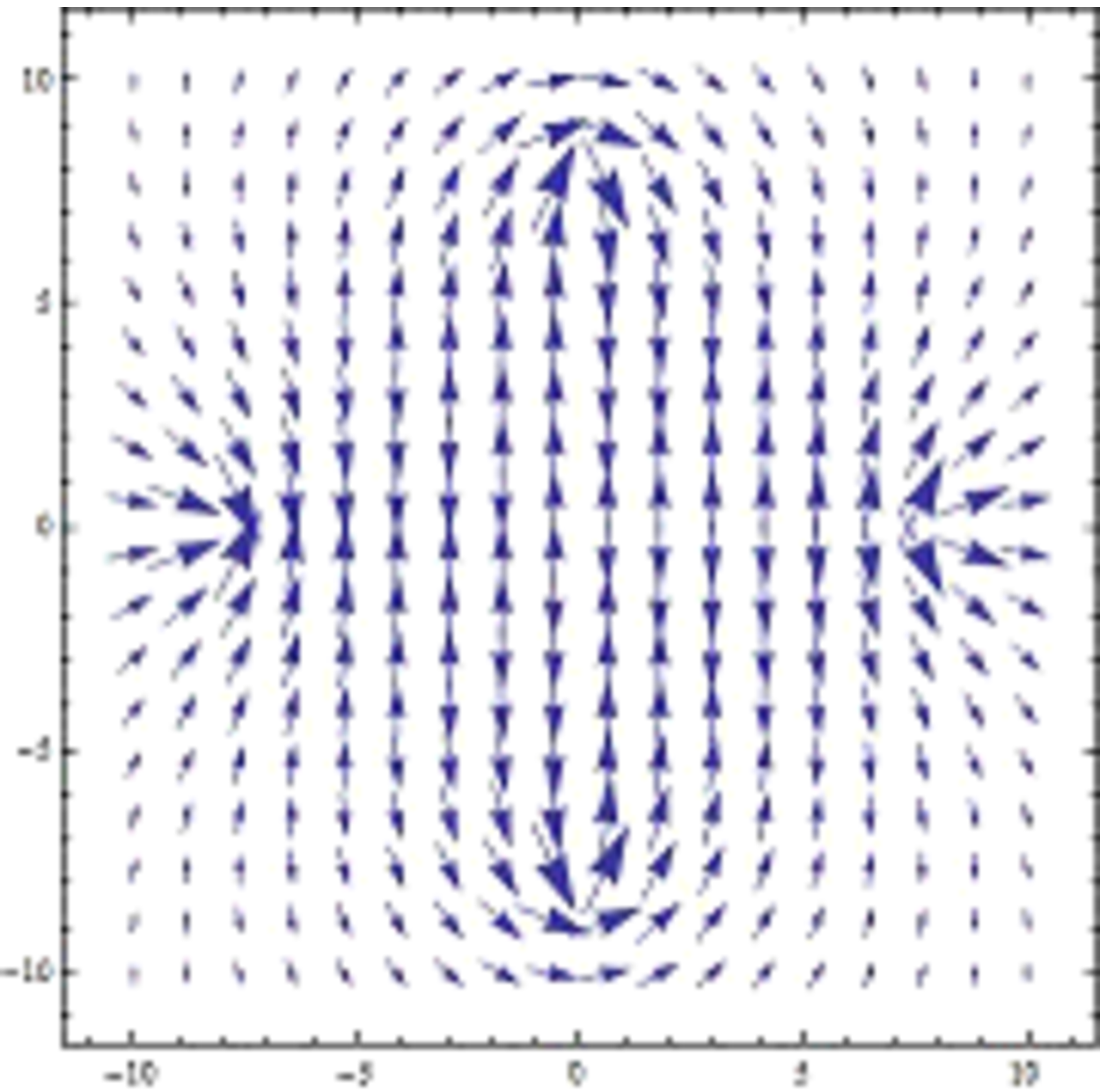}
\vspace{1cm}

\includegraphics[width=5cm]{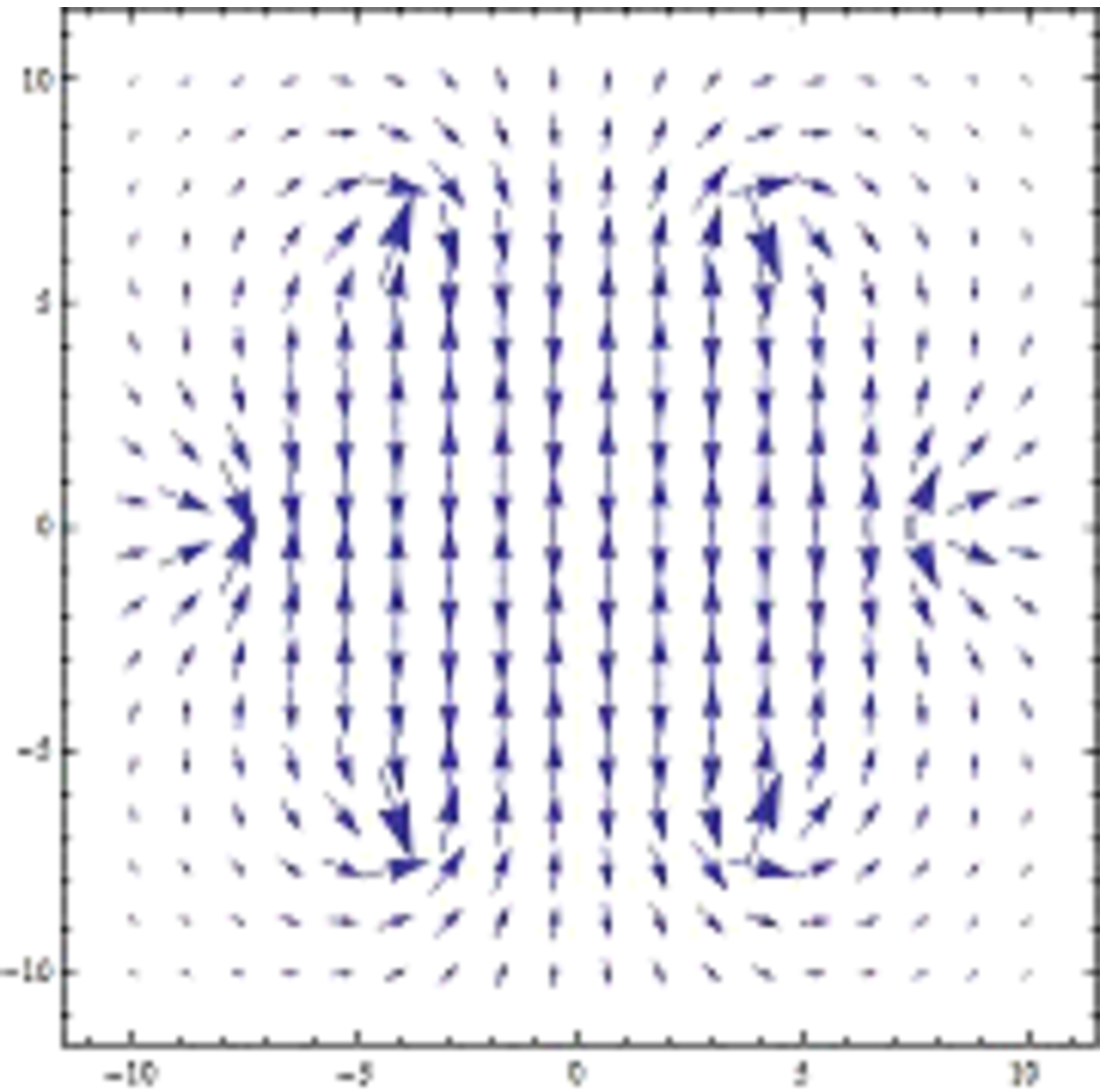}
 \caption{The arrow plot $(n_1,n_2)$ corresponding to the solution (\ref{sol_half}) 
of the $Q=2$ (top), $3$ (bottom) with $\beta e^2=2.0$, $k=0.0$ and $c=2.0$.}
\label{hvarrowplot}
\end{figure}

\subsubsection{The isosceles triangle}
Is it possible to give further generalization of the scheme of the construction such as an isosceles triangle for $N=3$?
The schematic picture is shown in Fig.~\ref{isoscelestriangle}. Here the centers sit $(d_1,d_2), (d_1,-d_2)$ and $(-c,0)$ in
the cartessian coordinate.  
The relation between the $(\ell_i,q_i)$, the polar coordinate $(\rho,\varphi)$ and $(d_1,d_2)$ are
\begin{align}
\ell_1 &:=\sqrt{ c^2+2 c \rho  \cos\varphi+\rho ^2},\notag \\
q_1 &:=\arctan\biggr[ \frac{\rho  \sin\varphi}{c+\rho  \cos\varphi}\biggr]; \notag \\
\ell_2 &:=\sqrt{ (\rho  \cos \varphi-d_1)^2+(\rho  \sin\varphi-d_2)^2}, \notag \\
q_2 &:=\arctan\biggr[ \frac{\rho  \sin \varphi -d_2}{\rho  \cos \varphi -d_2}\biggr]; \notag \\
\ell_3 &:=\sqrt{ (\rho  \cos \varphi-d_1 )^2+(\rho  \sin \varphi+d_2)^2},\notag  \\
q_3 &:=\arctan\biggr[ \frac{\rho  \sin \varphi +d_2}{\rho  \cos \varphi -d_2}\biggr].
\label{coordinatesrelationFacet}
\end{align}
Finally, we write down the form of the solution as
\begin{align}
u&=\Bigl(\frac{\ell_1}{a}\Bigr) \Bigl(\frac{\ell_2}{a}\Bigr) \Bigl(\frac{\ell_3}{a}\Bigr)e^{i[(q_1+q_2+q_3)+k(t+z)]}\notag\\
 &=\left(\rho  e^{i \varphi}+c\right) \left(\left(\rho  e^{i \varphi }-\tilde{d_1}\right)^2+\tilde{d_2}^{\hspace{3pt}2}\right),
\label{sol_facet}
\end{align}
where $\tilde{d_1}=d_1/a,\tilde{d_2}=d_2/a$. Again we omit `~$\tilde{~}$~' .
Note that the solution (\ref{sol_facet}) is holomorphic and then, it satisfies the zero curvature condition (\ref{zccondition}).
Substituting (\ref{sol_facet}) into (\ref{difeq_pot}) and a guess of suitable form of 
the potential, the form is determined such as
\begin{align}
V_{\tilde{3}}=& -\frac{16 \left(\beta  e^2-1\right)}{e^2 \left(1+(\mathcal{A}+c) (\mathcal{A}^*+c)\mathcal{B}\right)^4} \notag \\
&\times\left((d_1-\mathcal{A}) (d_1-3\mathcal{A}-2 c)+d_2^2\right)^2 \notag\\
&\times\left((d_1-\mathcal{A}^*) (d_1-3\mathcal{A}^*-2 c)+d_2^2\right)^2,
\label{pot_facet}
\end{align}
where $\mathcal{A}$ and $\mathcal{B}$ are defined as
\begin{align}
&\mathcal{A}=  \frac{1}{3} (2 d_1-c), \notag \\
&+\frac{\left(\sqrt{\zeta ^2+4 \tau ^3}+\zeta \right)^{\frac{1}{3}}}{3 \sqrt[3]{2}}
-\frac{\sqrt[3]{2} \tau }{3\left(\sqrt{\zeta ^2+4 \tau ^3}+\zeta \right)^{\frac{1}{3}}},\notag \\
&\mathcal{B}=\left((d_1-\mathcal{A})^2+d_2^2\right) \left((d_1-\mathcal{A}^*)^2+d_2^2\right),\notag \\ 
&\zeta= 27 u -18 d_2^2 (c+ d_1)-2 (c+d_1)^3,\notag \\
&\tau= 3 d_2^2-(c+d_1)^2.
\end{align}
In the special choice $d_1=\frac{c}{2}$, 
the potential (\ref{pot_facet}) perfectly agrees with (\ref{pot_3center}).

\begin{figure*}[t]
  \includegraphics[width=6cm]{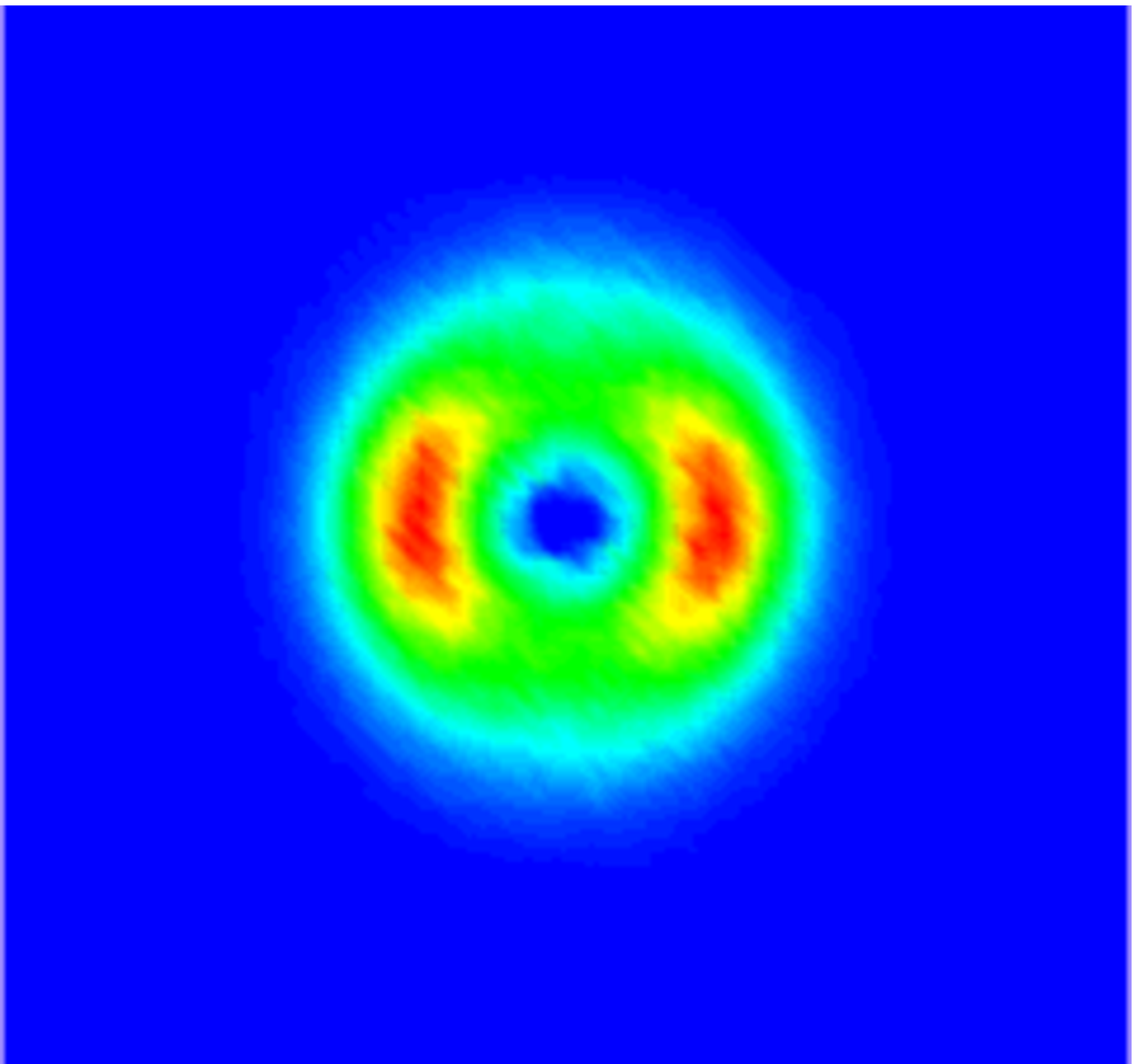}
\includegraphics[width=6cm]{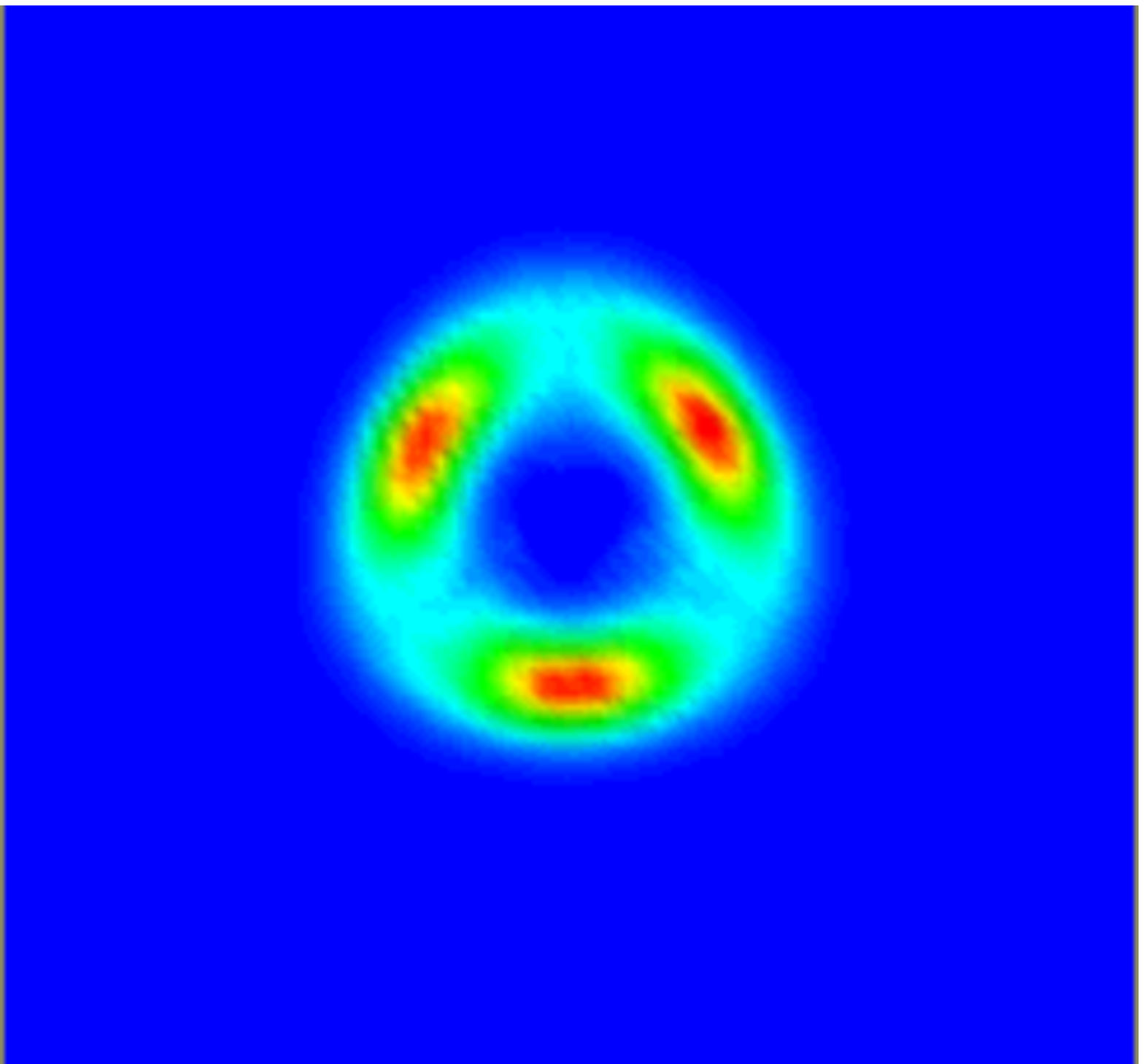}\\
 \includegraphics[width=6cm]{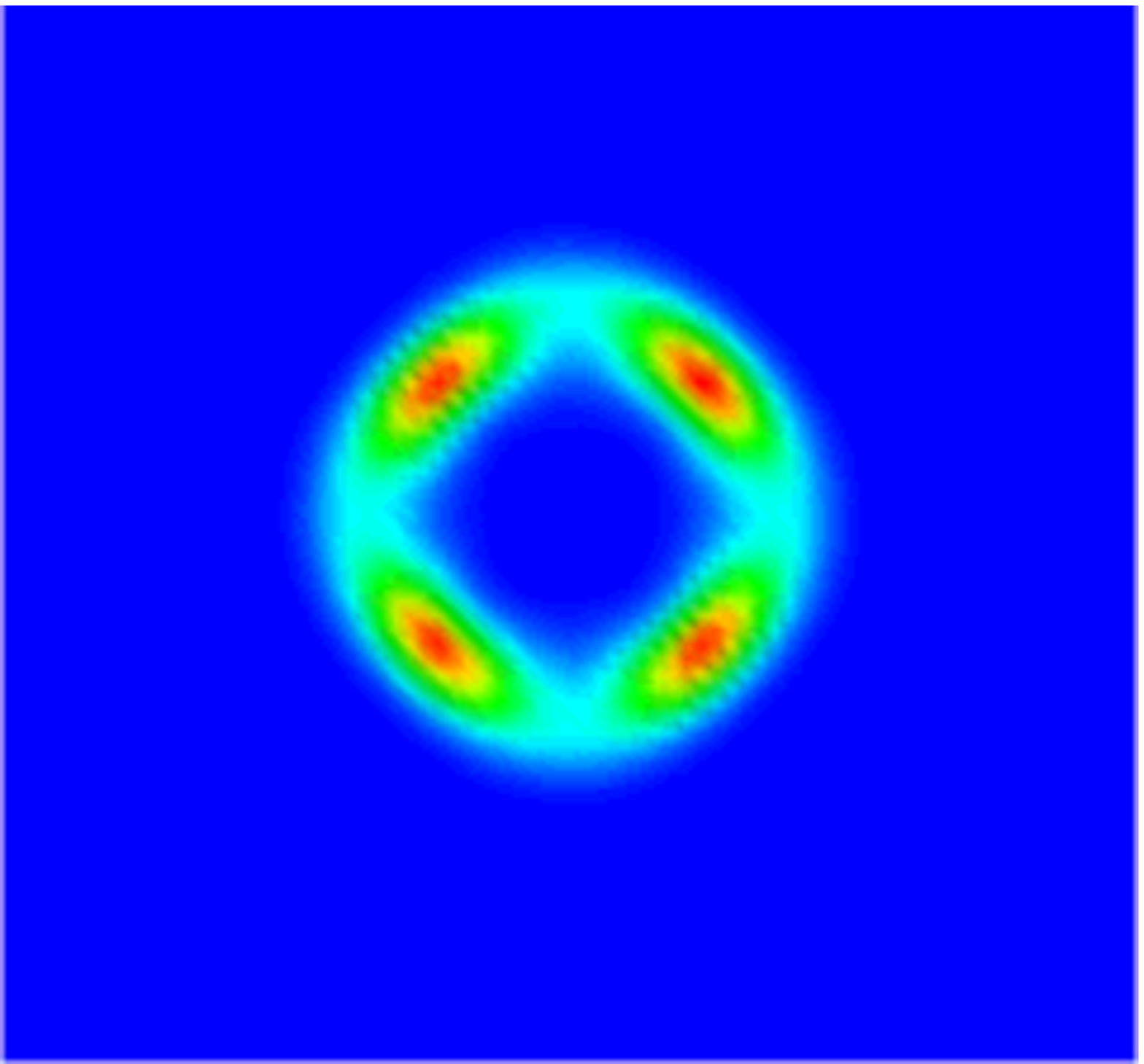}
\includegraphics[width=6cm]{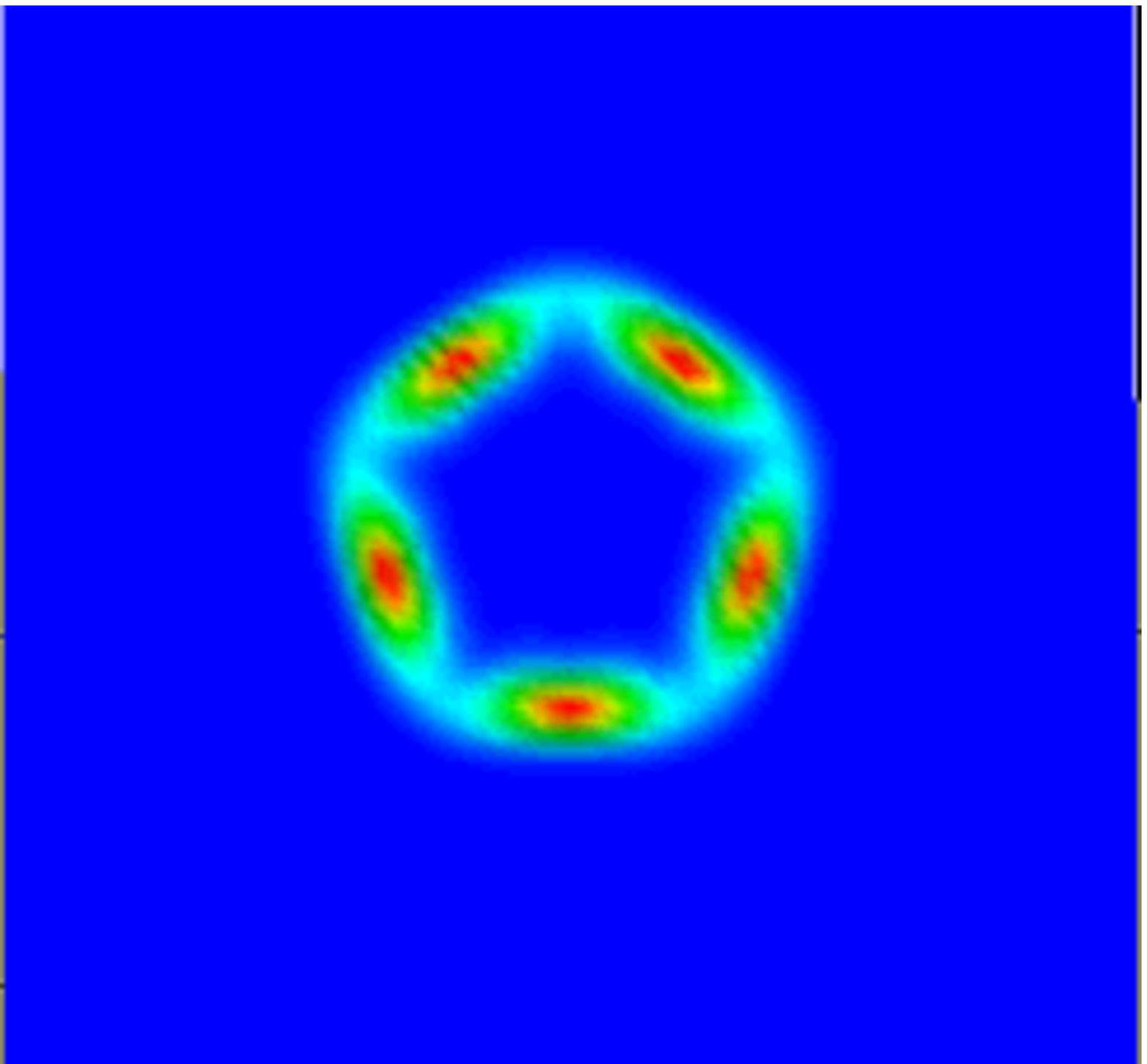}
 \caption{The hamiltonian densities of the $Q=2,3,4,5$ for  
$\beta=-2.0, e^2=-1.0, k=0.0$ and $c=1.0$.}
\label{energydensity}
\end{figure*}

\subsubsection{The half-integer charged vortex solution and the potential}

In \cite{Nitta}, a fractional vortex polygons are studied in terms of  
a simple potential $V=n_3^2$ which is inspired in antiferromagnets and the $XY$ model.
Our scheme can apply for construction of the half-integer charged vortex solution and the corresponding potential.
The solution  can be written as an extension of (\ref{sol_ncenter}) 
\begin{align}
u=\sqrt{\rho^{\hspace{3pt}N} e^{i[N\varphi+k(t+z)]}-c^{\hspace{3pt}N} e^{i[N\alpha+k(t+z)]}}.
\label{sol_half}
\end{align} 
It is almost straightforward to check that (\ref{sol_half}) satisfies the zero curvature condition (\ref{zccondition}).

Substituting (\ref{sol_half}) into (\ref{charge}), one can see that the topological charge $Q=\frac{N}{2}$. Then, 
(\ref{sol_half}) comprises the $N-$centered solution in which a each constituent carries half integer fraction of the charge. 
In order to see the structure in more detail, we present the arrow plot of $(n_1,n_2)$ for $Q=2,3$. 
One can easily see that the solutions exhibit a half of winding with each vortex core. 
Substituting (\ref{sol_half}) into (\ref{difeq_pot}), we have the potential of the form
\begin{align}
V_{\frac{N}{2}}=&\frac{\lambda}{256}\Big\{\left(n_1+in_2\right)^2+c^{\hspace{3pt}N} 
e^{i[N\alpha+k(t+z)]}(1-n_3)^2\Big\}^{2-\frac{2}{N}} \notag \\ 
&\times \Big\{\left(n_1-in_2\right)^2+c^{\hspace{3pt}N} e^{-i[N\alpha+k(t+z)]}
(1-n_3)^2\Big\}^{2-\frac{2}{N}} \notag \\
&\hspace{5cm}\times\frac{(1-n_3)^{\frac{8}{N}}}{(1-n_3^2)^2}.
\label{pot_half}
\end{align}

\begin{figure*}[t]
\includegraphics[width=15cm]{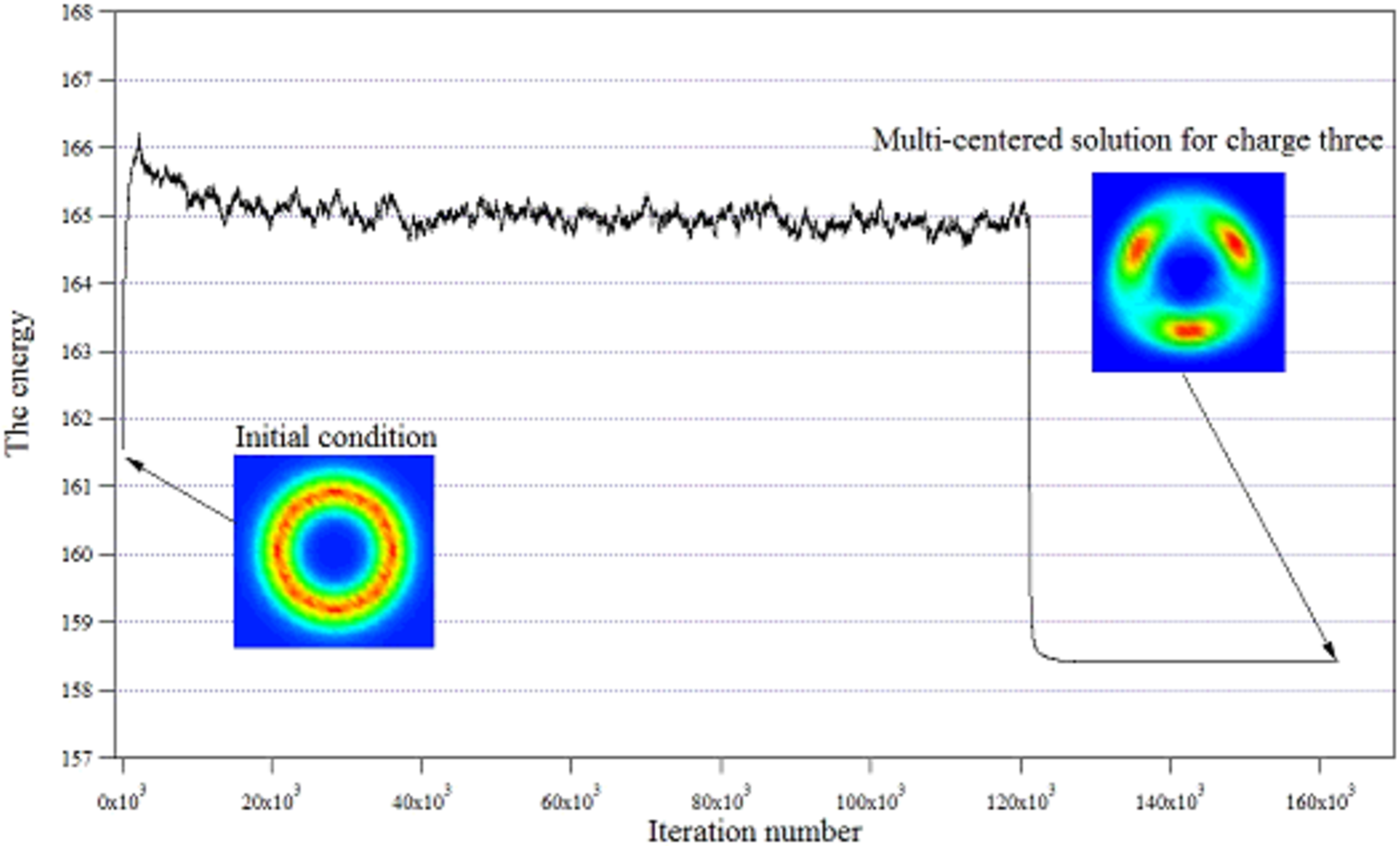}
 \caption{Energy per charge for the simulation of the $Q=3$ with $\beta e^2=2.0, e^2=-1.0, k=0.0$ and $c=1.0$.}
  \label{energy_iteration}
 \end{figure*}

\section{\label{sec:integrable}The zero curvature condition}
The potentials that we found in the previous section may be obtained in a slightly different way. 
Substituting the solution (\ref{sol_ncenter}) into the potential (\ref{potential_ex}), we get 
\begin{align}
V_N\to \frac{-32N^4(\beta e^2-1)\rho^{4N-4}}{a^{4N}\left(1+|u|^2\right)^4}.
\label{potential_Nsol}
\end{align}
On the other hand, substituting the solution (\ref{sol_ncenter}) into 
the second term of order four in derivatives of the fields
of the lagrangian (\ref{lagrangian2}), we get
\begin{align}
\frac{(\beta e^2-1)(\partial_{\mu}u\partial^{\mu}u^*)^2}{(1+|u|^2)^4} \to
\frac{4N^4(\beta e^2-1)\rho^{4N-4}}{a^{4N}\left(1+|u|^2\right)^4}.
\label{derivative_Nsol}
\end{align}
Apparently this means that at least on the holomorphic solutions, 
the potential is identified to the term of order four in derivatives of the fields 
up to some constants.
Thus we expect that when the solution is the holomorphic ones, 
the potential (\ref{potential_ex}) can always be rewritten as 
\begin{align}
V_N\to -8\frac{(\beta e^2-1)(\partial_{\mu}u\partial^{\mu}u^*)^2}{(1+|u|^2)^4}.
\label{potential_ex2}
\end{align}
By our zero curvature solutions, the term in the denominator of (\ref{potential_ex2}) 
generally can be calculated such as $\partial_{\mu}u\partial^{\mu}u^*=F(u,u^*)$, 
where the $F(u,u^*)$ is a function depends only on $u$ and $u^*$ 
(not on the derivatives).
If one plugs the explicit form of $F(u,u^*)$ into (\ref{potential_ex2}), one can 
obtain all potentials found in the previous subsections. 
Also, from (\ref{potential_ex2})
the potential is essentially equivalent to the term of order four in derivatives
of the fields, 
and it is straightforward to see that the current is conserved in terms of the
equivalence.

The zero curvature condition was first proposed in the context of the $CP^1$ model 
for integrable theories in any dimension \cite{Alvarez:1997ma},
and then applied to many models with target space being the sphere 
$S_2$, or $CP^N$~\cite{Alvarez:2009dt}. 
It leads to an infinite number of local conserved currents. 
Indeed, the equation of motion (\ref{sfeqn}) implies the conserved currents given by
\begin{align}
J^G_{\mu} \equiv \tilde{\mathcal{K}}_{\mu}\frac{\delta G}{\delta u}
-\tilde{\mathcal{K}}^{*}_{\mu}\frac{\delta G}{\delta u^{*}},
\label{currentG}
\end{align}
where $G$ is assumed to be any functional of $u,u^*$.
$\tilde{\mathcal{K}}_{\mu}$ can be defined as follows
\begin{eqnarray}
&&\tilde{\mathcal{K}}_{\mu} = \mathcal{M}^2 \partial_{\mu}u + \frac{4}{e^2}
\biggl[\frac{(\partial_{\nu}u\partial^{\nu}u)\partial_{\mu}u^*}{(1+|u|^2)^2} \notag \\ 
&&\hspace{0.5cm}+ \frac{2(\beta e^2 -1)(\partial_{\nu}u\partial^{\nu}u^*)\partial_{\mu}u}{(1+|u|^2)^2} \biggr]
\end{eqnarray}
in terms of the equivalence (\ref{potential_ex2}).

The current is conserved because in the derivative of the current
\begin{align}
\partial^{\mu} J^G_{\mu}&=\frac{\delta^2 G}{\delta u^2}\partial^{\mu}u\tilde{\mathcal{K}}_{\mu}
+\frac{\delta G}{\delta u}\partial^{\mu}\tilde{\mathcal{K}}_{\mu}+\frac{\delta^2 G}{\delta u \delta u^{*}}
\partial^{\mu}u^{*}\tilde{\mathcal{K}}_{\mu} \notag \\ 
&-\frac{\delta^2 G}{\delta u^{*2}}\partial^{\mu}u^{*}\tilde{\mathcal{K}}^{*}_{\mu}
-\frac{\delta G}{\delta u^{*}}\partial^{\mu}\tilde{\mathcal{K}}^{*}_{\mu}-\frac{\delta^2 G}{\delta u^{*}
 \delta u}\partial^{\mu}u\tilde{\mathcal{K}}^{*}_{\mu}.
\label{concerve}
\end{align}
The first and fourth term vanish due to the constraint (\ref{zccondition}),
and the third and sixth term cancel out, and 
the second and fifth term vanish because now the Euler-Lagrange equation has the form
\begin{eqnarray}
\partial^\mu\tilde{\mathcal{K}}_\mu=0,~~\partial^\mu\tilde{\mathcal{K}}_\mu^*=0.
\end{eqnarray} 
Thus the current $J^G_{\mu}$ is conserved and leads to an infinite number of conserved quantities.
If we choose the functional as 
\begin{eqnarray}
G:=-\frac{4i}{1+|u|^2},
\label{currentg1}
\end{eqnarray} 
one gets the corresponding conserved current
\begin{align}
J_{\mu}|_{U(1)}=&-4iM^2\frac{u \partial_{\mu}u^* -u^*\partial_{\mu}u}{(1+|u|^2)^2}  \\ \notag
&-i\frac{8(\beta e^2-1)}{e^2}\frac{2(\partial_{\nu}u\partial^{\nu}u^*)(\partial_{\mu}u^* u-u^* \partial_{\mu}u)}{(1+|u|^2)^4}
\end{align}
which corresponds to the Noether current for the (\ref{lagrangian}) associated to the symmetry $u \to u e^{i\alpha}$.
In Fig.~\ref{currentdensity}, we plot the components $J_t,J_{\rho},J_{\varphi}, J_{z}$ of the current 
for $N=3$.
Of course we are able to compute the current corresponding to the other symmetries. 
As an example, we plot a current in terms of 
\begin{eqnarray}
G:=-\frac{4 i|u|^2}{1+|u|^2}
\label{currentg2}
\end{eqnarray}
in Fig.~\ref{currentdensity2}.

\begin{figure*}[t]
\includegraphics[width=12cm]{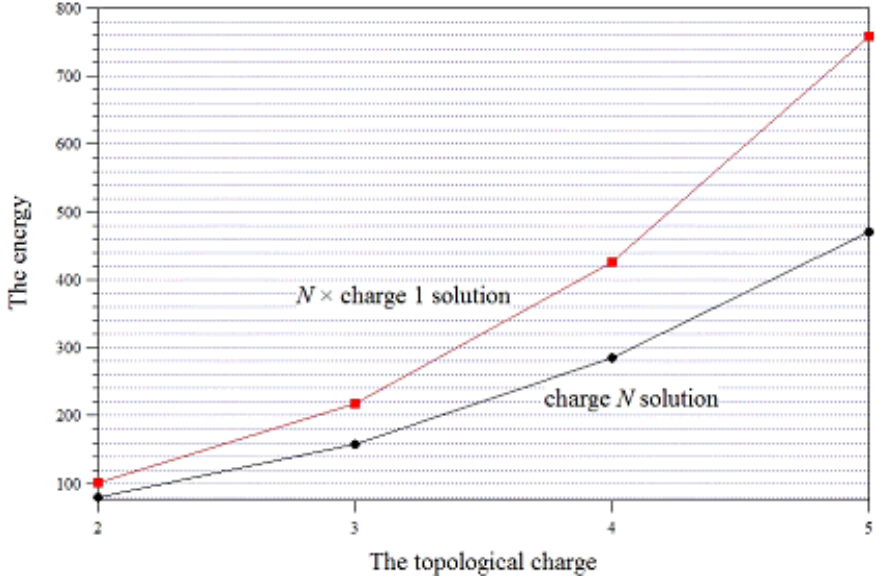}
 \caption{Energies of the solutions $Q=N=2,3,4,5$ and the ($N$ times of) the charge 1 
solution with $\beta e^2=2.0, e^2=-1.0, k=0.0$, and $c=1.0$.}
  \label{energy}
 \end{figure*}

\section{\label{sec:numerics}The numerical study} 
Assuming the form of the solutions which satisfy the zero curvature condition, 
we have obtained the potentials (\ref{pot_ncenter}) for general $N$.
It seems not straightforward that the converse is also true, 
i.e., such potentials can produce solutions which we initially assumed.
It is not easy question to answer and we solve the problem numerically.  
The simulated annealing method~\cite{Hale:2000fk} is a Hamiltonian minimization scheme which successfully 
finds the solution without any assumption for the solution.
However, in order to get solutions with infinite number of conserved quantities, a few conditions are 
applied to our numerical analysis. 
First, we assume the solution can be decomposed such as
\begin{eqnarray}
u={\cal U}(x,y){\cal W}(z,t)
\label{ansatz1}
\end{eqnarray}
where ${\cal U}, {\cal W} \in \mathbb{C}$. 
Second, we explore the solution traveling along $z$ axis with speed of light, which is consistent with our analytical solutions.  
If ${\cal W}$ is a function of the light-cone coordinate, i.e.,
${\cal W}(z,t)\equiv W(z\pm t)$,
one can easily see that the lagrangian (\ref{lagrangian2}) except for 
the potential term becomes static. In general, the potential still has some 
time dependence. It also becomes static when we consider the case
\begin{eqnarray}
{\cal W}(z, t) \equiv e^{ik(z\pm t)}.
\label{ansatz2}
\end{eqnarray}
In order to get solutions with infinite number of 
conserved quantities, the crucial point is that 
the potential can be rewritten as one of the term of order four in derivatives of the field 
as we discussed in the last section. 
Thus, if the fourth order term is static, the potential should also be time independent. 
This clearly indicates that the ansatz (\ref{ansatz1}) and (\ref{ansatz2}) are valid for finding 
solutions with infinite number of conserved quantities. 
As a result, we explore solution by minimizing the static Hamiltonian based on the ansatz
(\ref{ansatz1}) and (\ref{ansatz2}).  
Of course, we admit that outside these conditions there might be some solutions 
which still satisfy the zero curvature condition, but finding such solutions is 
obviously challenging but has to be quite difficult task. 
Thus we concentrate on analysis based on our assumption. 
If one evaluates the Hamiltonian density $\cal H$ for the Lagrangian (\ref{lagrangian2}) and then imposes the
conditions (\ref{ansatz1}) and (\ref{ansatz2}) one obtains

\begin{align}
{\cal H}=&\frac{4{\mathcal M}^2}{(1+|{\cal U}|^2)^2}(2k^2|{\cal U}|^2+\partial_x {\cal U} \partial_x {\cal U}^*+\partial_y {\cal U} \partial_y {\cal U}^*)\notag \\
&+\frac{8}{e^2(1+|{\cal U}|^2)^4}\big\{\big((\partial_x{\cal U})^2+(\partial_y{\cal U})^2\big) \notag \\
&\times\big((\partial_x{\cal U}^*)^2+(\partial_y{\cal U}^*)^2\big)-2k^2{\cal U}^2\big((\partial_x{\cal U}^*)^2+(\partial_y{\cal U}^*)^2\big)\notag \\
&-2k^2{\cal U}^{*2}\big((\partial_x{\cal U})^2+(\partial_y{\cal U})^2\big)\big\}\notag \\
&+\frac{8(\beta e^2-1)}{e^2(1+|{\cal U}|^2)^4}\big\{(\partial_x {\cal U} \partial_x {\cal U}^*+\partial_y {\cal U} \partial_y {\cal U}^*)\notag \\
&\times \big(4k^2|{\cal U}|^2+(\partial_x {\cal U} \partial_x {\cal U}^*+\partial_y {\cal U} \partial_y {\cal U}^*)\big)\big\}.
\label{HamiltonianCalU}
\end{align}

\begin{figure}[t]
  \includegraphics[width=6cm]{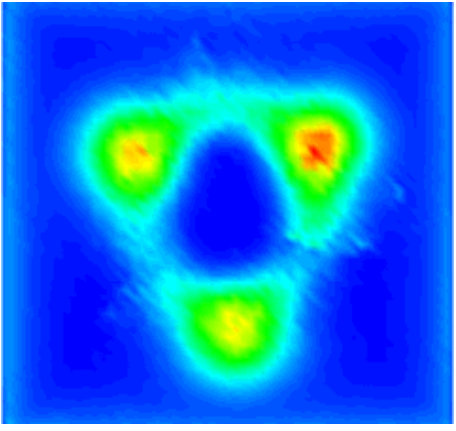}
 \caption{The hamiltonian density of the isosceles triangle
of the $Q=3$ with $\beta=-2.0, e^2=-1.0, k=0.0$ and $c=3.0$.}
\label{facet}
\end{figure}

\begin{figure*}[t]
  \includegraphics[width=12cm]{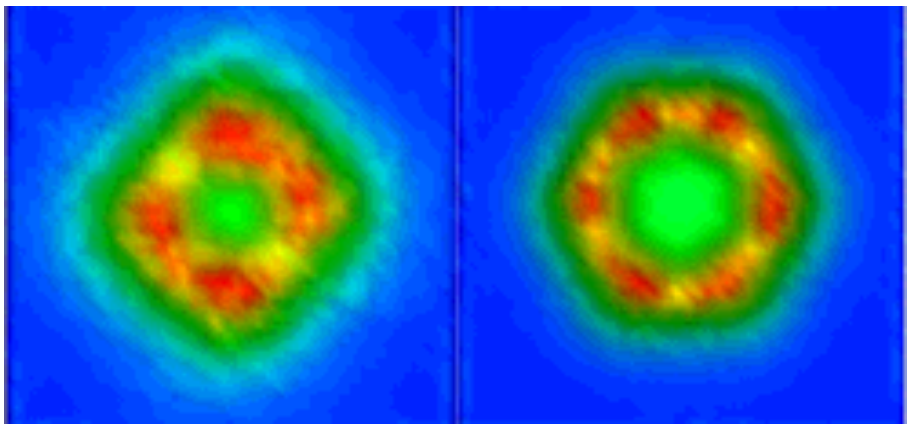}
 \caption{The hamiltonian density of the half vortex of the $Q=2$ (left), 3 (right) with $\beta=-2.0, e^2=-1.0, k=0.0$ and $c=1.0$.}
\label{halfvortex}
\end{figure*}

\begin{figure}[b]
  \includegraphics[width=10cm]{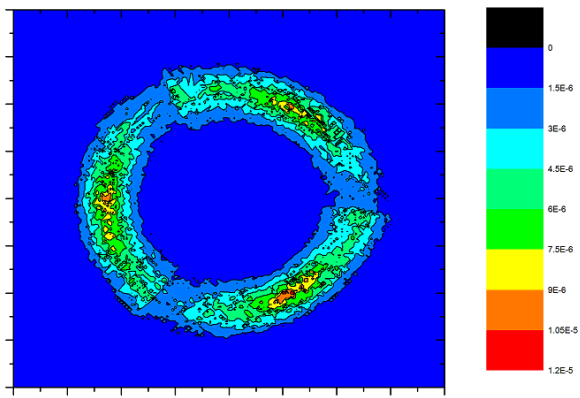}
 \caption{The zero curvature condition in terms of the annealing simulation, 
of the $Q=3$ with $\beta e^2=2.0, e^2=-1.0$, $k=0.0$ and $c=1.0$.}
  \label{zerocurv}
 \end{figure}

For ${\cal U}$, without loss of generality we can parameterize ${\cal U}:=F(x,y)e^{i\Theta(x,y)}$.
However, for the numerical simulation it is more convenient to use a static field $\vec{m}$ 
than ${\cal U}$ which has of the form
\begin{eqnarray}
\vec{m}
&=&
\( {\cal U}+{\cal U}^*,-i({\cal U}-{\cal U}^*),|{\cal U}|^2 -1\)/\( 1+|{\cal U}|^2\)
\nonumber\\
&=&\(\sin f\cos \Theta, 
\sin f\sin \Theta,\cos f \),
\label{ansatz_n}
\end{eqnarray}
where $\sin f(x,y):=\frac{2F(x,y)}{1+F(x,y)^2},\cos f(x,y):=\frac{F(x,y)^2-1}{1+F(x,y)^2}$. 

Substituting (\ref{ansatz_n}) into (\ref{HamiltonianCalU}) we have a form of static Hamiltonian density
\begin{align}
&{\cal H}_{x y}=\mathcal{M}^2\{{\cal P}_x+{\cal P}_y+2k^2(1+\cos^2f)\} \notag \\
&\hspace{0.5cm}- \frac{1}{2e^2}\bigg\{{\cal P}_x^2+{\cal P}_y^2
+2(f_xf_y+\sin^2f\Theta_x\Theta_y)^2 \notag \\
&\hspace{1cm}+2\sin^2f(f_x\Theta_y-f_y\Theta_x)^2
-2k^2\sin^2f({\cal Q}_x+{\cal Q}_y)\bigg\} \notag \\
&\hspace{0.5cm}- \frac{\beta e^2-1}{2e^2}({\cal P}_x+{\cal P}_y)\big({\cal P}_x+{\cal P}_y+4k^2\sin^2f\big) \notag \\
&\hspace{0.5cm}+V
\end{align}
where
\begin{align}
{\cal P}_{x,y}:=f_{x,y}^2+\sin^2f\Theta_{x,y}^2,~{\cal Q}_{x,y}:=f_{x,y}^2-\sin^2f\Theta_{x,y}^2
\end{align}
with $f_x=\frac{\partial f}{\partial x}, f_y=\frac{\partial f}{\partial y}, 
\Theta_x=\frac{\partial \Theta}{\partial x}$, and $\Theta_x=\frac{\partial \Theta}{\partial y}$. 
We numerically minimize the static energy ${\cal H}_{xy}|_{k=0}$. 
The static potential, e.g., the case of the standard integer $N$, 
can be written as
\begin{align}
V_{N}&=\frac{\lambda}{16}\Big\{\sin{f}e^{i\Theta}+c^{\hspace{3pt}N} 
e^{iN\alpha}(1-\cos{f})\Big\}^{2-\frac{2}{N}} \notag \\ 
&\hspace{1cm}\times \Big\{\sin{f}e^{-i\Theta}+c^{\hspace{3pt}N} e^{-iN\alpha}
(1-\cos{f})\Big\}^{2-\frac{2}{N}} \notag \\
&\hspace{1cm}\times (1-\cos{f})^{\frac{4}{N}}.
\label{pot_static}
\end{align}
The simulated annealing is a Monte-Carlo simulation in which one improve 
the value of the fields $\vec{m}$ (or equivalently $f$ and $\Theta$) 
by using a random numbers so as reducing the energy.
However, a more sophisticated method should be applied to the present problem.  
The method is the application of the Metropolis algorithm 
which can successfully avoids the unwanted saddle points.

\begin{figure*}[t]
  \includegraphics[width=6cm]{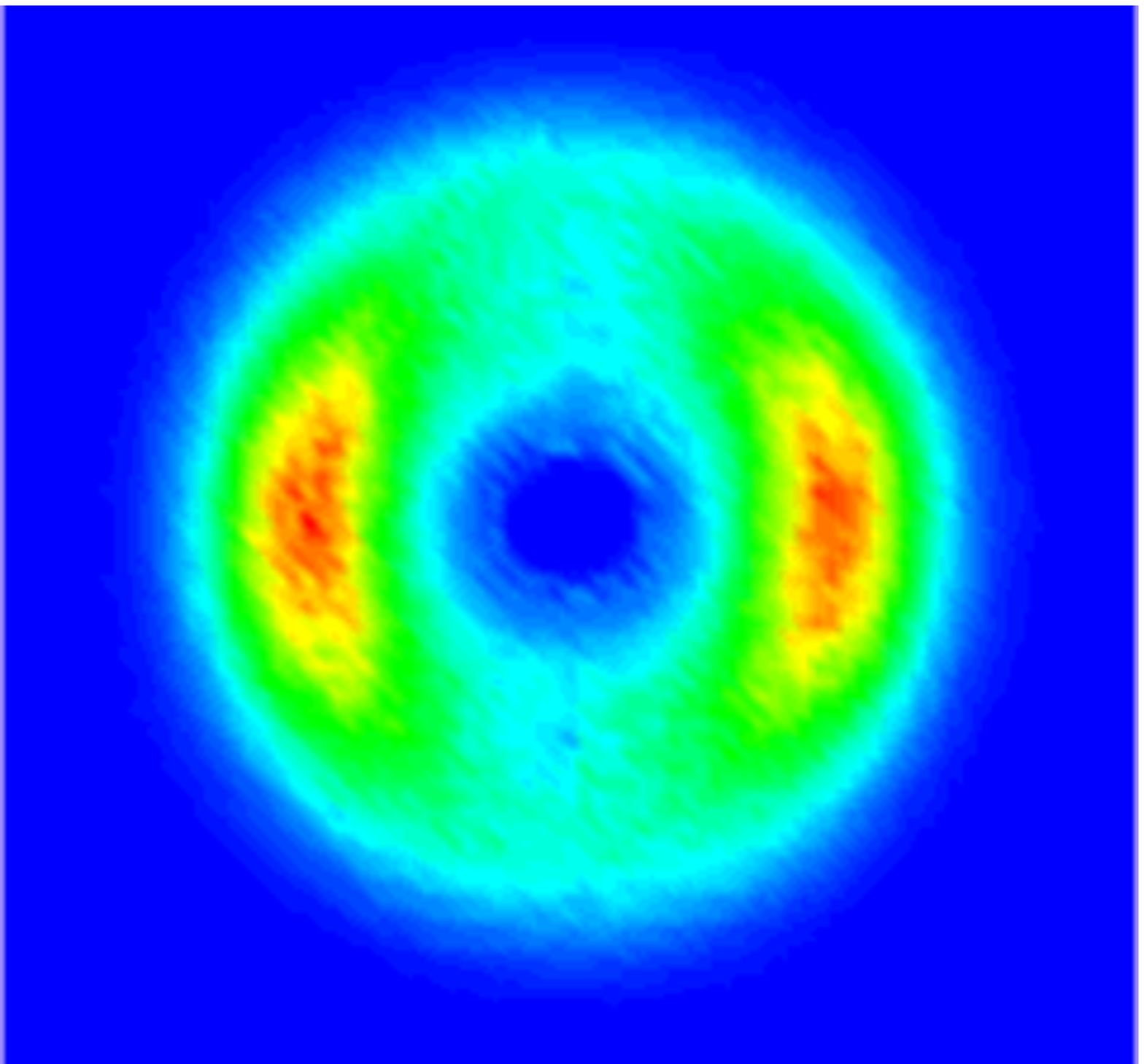}
\includegraphics[width=6cm]{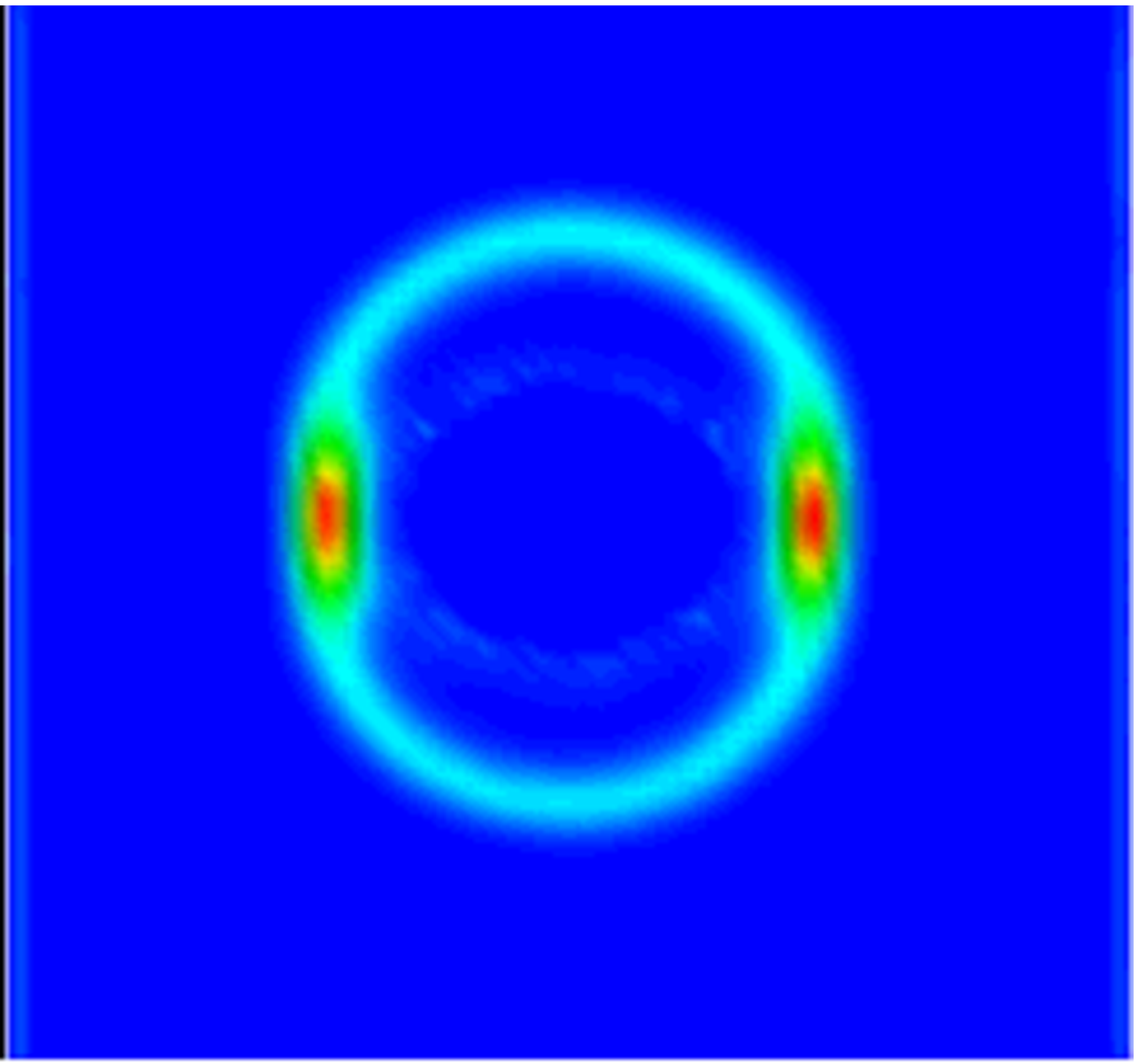}\\
 \caption{The hamiltonian densities with $k=0$ (left), $0.02$ (right) 
of the $Q=2$ with $\beta=-2.0, e^2=-1.0$,and $c=1.0$.}
\label{travelingwave}
\end{figure*}

The actual computation is done on the plane ${\cal P}:=(-d_x\leqq x\leqq d_x,-d_y\leqq y\leqq d_y)$ 
of which $d_x,d_y$ is a suitable size of the plane. We choose the mesh number 
$(N_x,N_y)=(80,80)$ for a good convergence.  
We impose the boundary conditions 
\begin{align}
&f(1,j)=\pi,\hspace{0.8cm}f(i,1)=\pi,\notag \\
&f(N_x,j)=\pi, \hspace{0.5cm}f(i,N_y)=\pi, \notag \\
&\Theta(1,j)=\pi/2,\hspace{0.4cm}\Theta(i,1)=\pi/2,\notag \\
&\Theta(N_x,j)=\pi/2, ~\Theta(i,N_y)=\pi/2 
\label{boundarycondition}
\end{align}
where $1 \leqq i \leqq N_x,~1 \leqq j \leqq N_y$.

In Fig.\ref{energydensity}, we present numerical results of the static hamiltonian density of $N=2,3,4,5$.
The solutions exhibit number $N$-peaks of which each have unit topological charge. 
In Fig.~\ref{energy_iteration}, we show the detail of how the simulation attained to the final answer. 
We plot the total energy of $N=3$ for the time step of the simulation. 
We start with the lump like $N=3$ solution as the initial profile, and after a certain period  
of heating we successfully reach the three-centered solution which has lower energy than the initial one. 
They do not depend on the choice of the initial profile and then we conclude the $N$-centered solution is 
ground state within the choice of potential $V_N$.

Fig.~\ref{energy} shows 
the total energies of $N=2,3,4,5$ compared with the energy of ($N$ times of) $N=1$ solutions which
are also computed using the potential $V_N$. 
Since the energy of the $N-$centered solutions are much lower than the ($N$ times of) $N=1$ solution, 
the solutions do not split into $N$ independent fractions. 

Next we present result of the isosceles triangle. In this simulation, 
we begin with a $N=3$ solution of which the centers are located on a straight line, 
and after a certain period of heating we successfully reach the isosceles triangle solution,  which of course 
has lower energy than the initial one. 

In Fig.~\ref{facet} we plot the static hamiltonian density ${\cal H}_{xy}|_{k=0}$ with the potential (\ref{pot_facet}) 
with $d_1=c/\sqrt{2}$.
Finally, we show the case of  half-integer charged vortex.
In Fig.~\ref{halfvortex}, we present numerical results of the hamiltonian density of $Q=2,3$.

As we have shown that our analytical solutions satisfy the zero curvature condition and then they 
possess infinite number of conserved quantities.  
So, it is quite interesting to check whether the numerical solutions also satisfy or not. 
In Fig.~\ref{zerocurv}, we plot the condition for our numerical solution for all 
the plane ${\cal P}$. Apparently the value is not zero for every point in ${\cal P}$. 
Note however that it mainly because of the size effect (choice of $d_x,d_y$) and the mesh discretization error
(which depend on the choice of both $d_x,d_y$ and the mesh number $N_x,N_y$). 
In fact, it is quite small than the net energy, 
i.e., the ratio is always about $\sim 10^{-8}$ which is order of the numerical uncertainty.

All the simulations shown in above are the case of $k=0$. 
For $k \neq 0$ it might be expected that the energy grows as $k$ increases, but it is not true. 
In Fig.~\ref{travelingwave}, we present results of the Hamiltonian density for $k=0$ (left) and $k=0.02$ (right).
The density suddenly becomes thin when we take finite $k$ and then, the solution finally becomes unstable for 
larger $k$.
This behavior may be understood in terms of the Derrick's argument. The stability of our vortex essentially should be
discussed in planar space, i.e., the equality holds between the quartic and the potential terms. The quartic term 
of the Hamiltonian contains the terms with $k^2$ while if we employ (\ref{ansatz2}), the potential does not.
Then, for increasing $k^2$ the equality is reached by decreasing the contribution of the terms which do not 
contain $k^2$. Therefore we have to conclude that the solution proposed in \cite{vortexlaf} becomes unstable if
$\beta e^2\neq 1$, while the static solution $k=0$ certainly exists. Of course outside the assumption (\ref{ansatz2})
the solution might exist because the potential may have time dependence. 
The study certainly is challenging because the simulation with all the coordinates including $z,t$ is required. 
(Such kinds of solutions might eventually be unstable as already shown in \cite{Hietarinta:2003vn}.) 
We will report the results in a future article. 

\section{\label{sec:summary}Summary and Outlook}
In this paper, we mainly discussed how to get analytical, multi-centered vortex solutions 
and corresponding potentials of the extended version of the Skyrme-Faddeev model. 
We found forms of the potential (\ref{pot_ncenter}) for our ansatz of the $N$-centered solutions.
We confirmed that such potentials coincide with the previous studies for the one- and 
two-centered solutions \cite{Ferreira:2011mz, Piette:1992he}. 

There are number of studies to get such multi-centered solutions using several choices of potentials. 
Most of the studies are based on the numerical analysis, and the solutions are static solutions~\cite{Jaykka:2011ic,Nitta}. 
Contrary to those cases, we found the analytical, traveling wave vortex solutions going to the (minus) $z$ direction.  
Furthermore they have the infinite number of conserved quantities and then are in the integrable sector. 
Next, by examining the full-field relaxation method, 
we confirmed that the potentials perfectly work well to get such integrable solutions from arbitrary initial profile. 

Our scheme is quite general and is easily applicable to the related two dimensional 
solitonic models such as the baby-Skyrme model, the $CP^N$ sigma model and so on.
For the physical application, some vortex states in a superconductor may be possible candidates of our solutions. 
The model has a relationship with the standard electroweak theory, especially when one considers the case of 
a global $SU(2)$ and a local $U(1)$ breaking into a global $U(1)$, where 
the model reduces to an Abelian Higgs model with two charged scalar fields~\cite{Achucarro:1999it}. 
It is interesting to note that the vortices of such model carry the so-called longitudinal electromagnetic 
currents~\cite{Forgacs:2006pm,Volkov:2006ug}. 
Furthermore, the higher winding number solutions exhibit a pipe-like structure~\cite{Chernodub:2010sg}.

Our model enjoys a symmetry breaking of the type $O(3)_{\rm global} \to {D_N}_{\rm global}$ 
which is complicated than $SU(2)_{\rm global}\otimes U(1)_{\rm local}\to U(1)_{\rm global}$.
It is certainly interesting that we explore the conserved quantities and the nontrivial structures 
of a type II superconductors by our prediction. 

It is also worthwhile to apply our technique to more realistic three dimensional problems. 
The Skyrme model is a low energy effective model of nuclei and the multi-winding number solutions exhibit
the platonic symmetries~\cite{Braaten:1989rg,Battye:1997qq}. 
The main drawback of such solutions is that the density becomes zero at the vicinity of the center and it is inconsistent with structure of the actual nuclei. 
There is a certain possibility that our molecule type ansatz may cure of such difficulty and it may 
describe the basic properties of the observed nuclei, e.g., the charge density.

\noindent {\bf Acknowledgments} 
We are grateful to Luiz Agostinho Ferriera and Pawe\l~Klimas for many valuable discussions.
We also would like to thank to Wojtek Zakrzewski for useful comments.
We thank to Masahiro Hayasaka for his support for the numerical analysis.   
We also thank for the kind hospitality at Instituto de F\'isica
de S\~ao Carlos, Universidade de S\~ao Paulo.  
YT also acknowledges the financial support of Tokyo University of Science.

\end{document}